\documentclass[a4paper,twocolumn,11pt,amsmath,amssymb,dvipsnames,accepted=2024-10-29]{quantumarticle}
\pdfoutput=1
\usepackage[utf8]{inputenc}
\usepackage[english]{babel}
\usepackage[T1]{fontenc}
\usepackage{amsmath}
\usepackage{tikz}
\usepackage{lipsum}
\usepackage[numbers]{natbib}

\usepackage{graphicx}
\usepackage{dcolumn}
\usepackage{bm}
\usepackage{hyperref}
\usepackage{xcolor}
\usepackage{soul}
\usepackage[left]{lineno}
\begin{document}

\title{Anomalous Floquet Phases. A resonance phenomena}

\author{\'Alvaro~G\'omez-Le\'on}
\email{a.gomez.leon@csic.es}
\affiliation{Institute of Fundamental Physics IFF-CSIC, Calle Serrano 113b, 28006 Madrid, Spain}
\begin{abstract}
Floquet topological phases emerge when systems are periodically driven out-of-equilibrium. They gained attention due to their external control, which allows to simulate a wide variety of static systems by just tuning the external field in the high frequency regime. However, it was soon clear that their relevance goes beyond that, as for lower frequencies, anomalous phases without a static counterpart are present and the bulk-to-boundary correspondence can fail.
In this work we discuss the important role of resonances in Floquet phases. For that, we present a method to find analytical solutions when the frequency of the drive matches the band gap, extending the well-known high frequency analysis of Floquet systems.
With this formalism, we show that the topology of Floquet phases with resonances can be accurately captured in analytical terms.
We also find a bulk-to-boundary correspondence between the number of edge states in finite systems and a set of topological invariants in different frames of reference, which crucially do not explicitly involve the micromotion.
To illustrate our results, we periodically drive a SSH chain and a $\pi$-flux lattice, showing that our findings remain valid in various two-band systems and in different dimensions.
In addition, we notice that the competition between rotating and counter-rotating terms must be carefully treated when the undriven system is a semi-metal.
To conclude, we discuss the implications to experimental setups, including the direct detection of anomalous topological phases and the measurement of their invariants.
\end{abstract}
\maketitle
\tableofcontents
\section{Introduction}
Condensed matter has rapidly evolved during the last decades by incorporating concepts from many different areas of physics and mathematics.
A quite remarkable example of this is the field of topological systems, where ideas from topology and particle physics have been borrowed to explain the fundamental properties of certain materials, which could not be understood using the standard classification of phases of matter based on local order parameters~\cite{SachdevBook}.
Nowadays, this interplay between topology and physics has been extremely fruitful and extends beyond the frontiers of condensed matter. For example, it is used to predict robust states with applications in quantum computation~\cite{Kitaev_2001}, to explain large-scale climate phenomena~\cite{Delplace2017} or even certain properties of stars~\cite{leclerc_topological_2022}.

In this plethora of topological phases of matter, non-equilibrium systems play an interesting role. Initially, Floquet topological insulators captured wide attention for being topological systems with an external control~\cite{Oka-2009,lindner_floquet_2011,grushin_floquet_2014,Diaz2019,Fang2022,Fang2023}. However, it was soon realized that their non-equilibrium nature provides a richer topology than their equilibrium counterparts~\cite{kitagawa_topological_2010,Rudner_2013,gomez-leon_floquet-bloch_2013,gomez-leon_engineering_2014,Torres-2014}. 
For example, Floquet anomalous phases exhibit robust topological edge states with trivial band invariants, something that is utterly impossible in equilibrium systems.
Currently, topology in non-equilibrium systems not only involves periodically driven systems, but also dissipative ones~\cite{PhysRevA.106.L011501}, which have shown to share extraordinary connections with Floquet systems~\cite{bessho_nielsen-ninomiya_2021}.\\
On the experimental side, there has been a lot of progress on the observation of topological phenomena, and in particular, Floquet topological insulators have been realized on the surface of topological insulators~\cite{Gedik2013}, in cold atoms~\cite{wintersperger_realization_2020,Braun2024}, photonic crystals~\cite{Rechtsman2013,cheng_observation_2019,cheng_observation_2022} or optical fibers~\cite{upreti_topological_2020,adiyatullin_topological_2022}, among many others~\cite{mciver_light-induced_2019}.

Therefore, the study of Floquet phases is a timely research line and anomalous topology represents one of its most mysterious and interesting areas to explore.
Probably, part of this mystery is a consequence of the fundamental differences with static topological insulators and the sophisticated mathematical machinery typically used to characterize anomalous Floquet phases.
In particular, quasienergies in Floquet systems play a similar role as the energies of static systems, but they are $2\pi$-periodic. This leads to the existence of an additional unequivalent gap between quasienergies, known as $\pi$-gap.
Hence, instead of the $N-1$ gaps that are present in static systems with $N$ bands, Floquet systems with $N$ bands have $N$ unequivalent gaps.
This makes their topology more complex, which now involves new invariants that contain a micromotion, due to the possibility of band degeneracies during the time evolution~\cite{nathan_topological_2015,Nur2019,rudner_band_2020}.\\
From a more physical point of view, the consequences of these generalizations from static to nonequilibrium setups are not completely clear.
One clue to understand the physics behind the emergence of anomalous topology is given by the fact that it only exists out of the high frequency regime.
This indicates that the physics of resonances must play a pivotal role, and the transition between the high and the low frequency regime must be related to their existence~\cite{Benito2014,Dalibard2014,FloquetWannier}.

In this work, we study the importance of resonance phenomena in Floquet phases~\cite{Goldman2015,Perez-Piskunow2015} and, in particular, in the case of anomalous topology.
With this goal in mind, we describe a general procedure to find effective Hamiltonians that faithfully capture the physics of periodically driven systems in the presence of resonances. 
This approach generalizes well-established high frequency methods and unravels the interplay between resonant and off-resonant processes in the topology of non-equilibrium phases. Also, it accounts for large-amplitude modulations, as they play a fundamental role in Floquet engineering.

Our results show that resonances are a crucial ingredient in producing anomalous Floquet phases and that the phase boundary of the topological phase, as a function of frequency, can be accurately predicted in analytical terms. In addition, we discuss how the symmetries of the driving field affect the existence of phase transitions by forbidding or allowing exact crossings of quasienergies at particular points of the first Brillouin zone (FBZ).
Our results provide a solid methodology for studying real situations in experiments, where typically resonant and off-resonant states coexist.

Importantly, we find that the topology of Floquet phases, in the presence of resonances, can be expressed as the combined topology of different frames of reference. 
This allows to characterize the topology of Floquet phases without directly involving the micromotion, which is needed just to link the different frames of reference.
As a consequence, one can use the standard topological invariants over the FBZ, and establish a bulk-to-boundary correspondence between edge states and invariants in each frame of reference.

Finally, in the case of periodically driven semi-metals we also uncover a subtle interplay between rotating and counter-rotating terms. We show that they must be carefully treated to correctly capture both, the renormalization of the off-resonant states and the resonance phenomena.
\section{Revisiting the Rabi model}
The Rabi model is one of the most iconic models in quantum physics~\cite{Rabi1937}. It describes, at an elementary level, the interaction between radiation and quantum matter, and as such, it also is a canonical model for non-equilibrium physics.

Beyond its seemingly simple form, the Rabi model for arbitrary field does not have a closed form solution.
Despite this, different approximations can provide exact analytical solutions in different regimes of operation, and explain different phenomena such as Rabi oscillations or Floquet engineering.
These typically are the rotating wave approximation (RWA) or the high frequency expansions.

In this section we perform a basic analysis of the Rabi model for a two-level system coupled to a radiation field under the dipole approximation, and discuss its analytical solutions and their range of validity.
This will be useful to understand the limitations of our approximations in the case of lattice systems coupled to AC fields in the following sections.

The Hamiltonian has the following form:
\begin{equation}
    H\left(t\right)=\frac{\Delta}{2}\sigma_{z}+V\cos\left(\omega t+\varphi\right)\sigma_{x},\label{eq:Rabi1}
\end{equation}
being $\sigma_{x,y,z}$ the Pauli matrices that fulfill $\left[ \sigma_\mu, \sigma_\nu \right]= 2i\epsilon_{\mu\nu\eta} \sigma_\eta$, $\omega$ the frequency of the external field, $\varphi$ its initial phase, $V$ its amplitude and $\Delta$ the splitting of the two-level system (TLS). Importantly, notice that the interaction between the TLS and the field is transversal, which means that the field can drive transitions in the TLS.

In order to explore amplitudes beyond the weak coupling regime, defined by $V\ll\Delta,\omega$, we first perform a transformation to the interaction picture with
\begin{equation}
    R_{1}\left(t\right)=e^{-i \int V \cos\left(\omega t+\varphi \right) \sigma_{x} dt},
\end{equation}
which leads to a new, time-dependent Hamiltonian:
\begin{align}
H_{I}\left(t\right) =& R_{1}^{\dagger}\left(t\right)H\left(t\right)R_{1}\left(t\right)-iR_{1}^{\dagger}\left(t\right)\dot{R}_{1}\left(t\right)\nonumber\\
=& \frac{\Delta}{2}\cos\left[F(t)\right]\sigma_{z} +\frac{\Delta}{2}\sin\left[F(t)\right]\sigma_{y},\label{eq:HI1}
\end{align}
being $F(t)\equiv 2V \sin(\omega t+\varphi)/\omega$.
So far, this result is exact, but it is not clear why Eq.~\eqref{eq:HI1} is more convenient than Eq.~\eqref{eq:Rabi1}, as it is more complex and fully time-dependent.
The reason is that the field amplitude dependence is now encoded in the trigonometric functions. As they are bounded between $\pm 1$, this allows us to truncate the Hamiltonian for arbitrary field amplitude.
To do this, we just need to apply the well-known Jacobi-Anger expansion of the trigonometric functions in terms of Bessel functions:
\begin{align}
    H_{I}(t)=& \sum_{n=-\infty}^{\infty}\frac{\Delta_{n}}{2}\cos\left(n\omega t+n\varphi\right) \sigma_{z} \nonumber \\
&+\sum_{n=-\infty}^{\infty}\frac{\Delta_{n}}{2}\sin\left(n\omega t+n\varphi\right) 
\sigma_{y} \label{eq:HI2}
\end{align}
Notice that each of the Fourier components is renormalized by Bessel functions $\Delta_n=\Delta\mathcal{J}_{n}\left(2V/\omega\right)$, being $\mathcal{J}_n(z)$ the $n$-th Bessel function of the first kind. 
This is one of the simplest examples of Floquet engineering, where the parameters of the original Hamiltonian are externally tuned by the applied field~\cite{Gomez_Leon_2011,Gomez_Leon_2012,gomez-leon_floquet-bloch_2013,delplace_merging_2013}.
\begin{figure}
    \centering
    \includegraphics[width=\columnwidth]{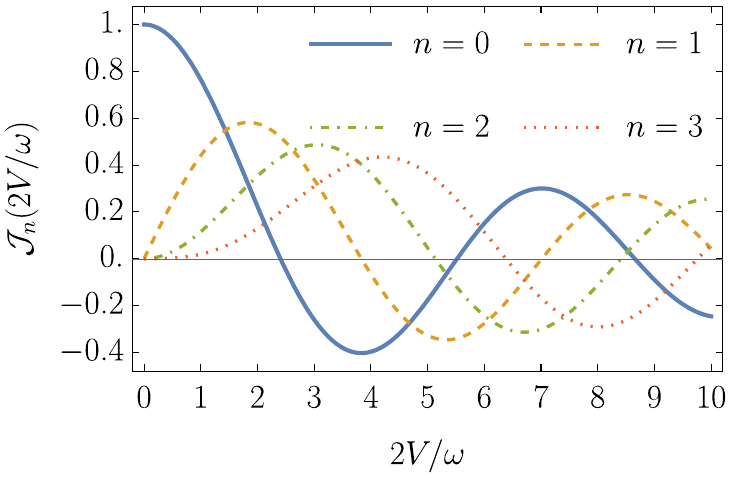}
    \caption{Bessel functions $\mathcal{J}_n(2V/\omega)$ for $n=0,1,2$ and $3$. They oscillate with slightly different periods and decay for increasing field amplitude.}
    \label{fig:Bessel0}
\end{figure}
Concretely, Eq.~\eqref{eq:HI2} is useful because generally, just a few Bessel functions dominate for a particular value of the ratio $V/\omega$.\\
Also, it is interesting to notice that the original, monochromatic driving in Eq.~\eqref{eq:Rabi1}, turns into a drive with an infinite number of harmonics in Eq.~\eqref{eq:HI2}, when the transformation to the interaction picture is applied. This is a way to understand how the complexity of strong field effects emerge in the form of high harmonics, once the transformation to the interaction picture is applied.\\

At this point it is useful to have some intuition about the the behavior of Bessel functions as the ratio $V/\omega$ increases. In Fig.~\ref{fig:Bessel0} one can see that the $n=0$ Bessel function dominates for small $V/\omega$. 
As it goes with the time-independent part of the Hamiltonian, this component can be interpreted as field renormalized bands, whose bandwidth tends to zero at large $V/\omega$. 
This band collapse at large $V/\omega$ is well-known, and shows that for very strong drive, bands tend to decouple, in some cases justifying the stroboscopic description of Floquet systems, even at intermediate frequencies.
Physically, it can be understood as a large energy shift that brings the system out of resonance. In addition, one can see that the zeroes of the Bessel function correspond to a phenomena known in the literature as Coherent Destruction of Tunneling~\cite{Grossmann_1991}.\\
The other Bessel functions, $\mathcal{J}_{n\neq0}(2V/\omega)$, appear with the time-dependent contributions of the effective Hamiltonian. Hence, they can be interpreted as drive terms with an upper-bound value, set by the maximum of the corresponding Bessel function.
Then, it will be reasonable to assume that the largest of all will control the dynamics, specially if its frequency, $n\omega$, is close to the renormalized splitting of the TLS, $\Delta_0$. Usually the effect of the other components can be accurately captured by means of perturbation theory in powers of $\Delta_{n\neq 0}$, unless the system is very close to resonance with one of them.\\

For our purposes, we will consider that the field amplitude is such that $|\Delta_{1}|>|\Delta_{n \geq 2}|$, although one can carry out a similar analysis for a different range of field amplitudes. 
This allows us to assume that the interaction picture Hamiltonian can be efficiently truncated to the following form:
\begin{equation}
    H_{I}(t) \approx \frac{ \Delta_{0}}{2} \sigma_{z}+\Delta_{1}\sin \left( \omega t + \varphi \right) \sigma_{y}. \label{eq:HI3}
\end{equation}
Equation~\eqref{eq:HI3} looks very similar to Eq.~\eqref{eq:Rabi1}, with the difference that $V$ is not explicitly present. Its effect is encoded in the renormalized splitting, controlled by the zeroth Bessel function, and in the renormalized coupling to the field, controlled by the first Bessel function (plus an additional $\pi/2$ phase difference in the driving term). 
Hence, as previously claimed, we can now work with a Hamiltonian similar to the original one, with the advantage that is valid beyond weak field amplitudes.

When the frequency is the dominant energy scale (i.e., in the high frequency regime, $\omega\gg\Delta,V$), the description of the system highly simplifies, because the second term in Eq.~\eqref{eq:HI3} can be neglected. The reason for this is that non-zero Fourier components of the Hamiltonian oscillate very rapidly and average to zero over time.
This allows to just focus on the stroboscopic dynamics, which is given by the renormalized splitting, $\Delta_0$.

On the contrary, if the frequency is of the order of the splitting, the time-dependent part of Eq.~\eqref{eq:HI3} cannot be neglected and the dynamics must be included. This is the regime of Rabi oscillations, where an accurate solution can be obtained by means of a RWA that results in:
\begin{equation}
    H_\text{RWA}\left(t\right) = \frac{\Delta_{0}}{2}\sigma_{z}+\frac{\Delta_{1}}{2}\left[e^{i(\omega t+\varphi)}\sigma_{-}+e^{-i(\omega t+\varphi)}\sigma_{+}\right].\label{eq:HRWA1}
\end{equation}
To obtain Eq.~\eqref{eq:HRWA1} we have neglected the components proportional to $e^{\pm i (\omega t + \varphi)}\sigma_{\pm}$, which are the counter-rotating terms that are ignored in the RWA.
This Hamiltonian can be exactly solved by performing a transformation to a rotating frame $R_2 (t)=e^{-i\frac{\omega}{2}t\sigma_{z}}$, where it becomes time-independent. This results in:
\begin{equation}
    \tilde{H}=\frac{\Delta_{0}-\omega}{2}\sigma_{z}+\frac{\Delta_{1}}{2}\left(e^{-i\varphi}\sigma_+ + e^{i\varphi}\sigma_-\right)\ . \label{eq:HRWA1-1}
\end{equation}
Now the system can be exactly solved by just finding the eigenvalues of $\tilde{H}$:
\begin{equation}
    \lambda_\pm = \pm \frac{1}{2}\sqrt{\left( \Delta_0 -\omega \right)^2 + \Delta_1^2},\label{eq:eigenvalues1}
\end{equation}
and the eigenvectors $| \phi_{\pm} \rangle$. The time evolution of the eigenstates can be expressed in the original reference frame as:
\begin{equation}
    |\Psi_{\pm}(t) \rangle = R_1(t) e^{-i\frac{\omega}{2} \sigma_{z} t} e^{-i\lambda_{\pm}t}| \phi_{\pm} \rangle \label{eq:eigenvectors1}
\end{equation}
Notice from Eq.~\eqref{eq:eigenvalues1} that, as the frequency decreases towards $\Delta_0$, the splitting gets reduced, and precisely at resonance there is an anticrossing controlled by the renormalized coupling strength $\Delta_1$. 
Only for a zero of the Bessel function $\mathcal{J}_{1}(2V/\omega)$ the spectrum could exactly closes its gap.
Importantly, the RWA is highly accurate at resonance $\omega=\Delta_0$, because resonances are non-perturbative phenomena that mask the perturbative physics underneath. This will be further discussed in the next sections.\\

Let us now express these results using the language of Floquet theory, as it will be important to link the quasienergies with the eigenvalues in different frames~\cite{Benito2014}.\\
Floquet theory can be applied to any system with a periodic time-dependence~\cite{GRIFONI1998}. In analogy with spatially periodic systems and Bloch theorem, it states that the solutions can be written in terms of conserved quantities called quasienergies $\epsilon_\alpha$, and Floquet states $|\Phi_\alpha (t) \rangle$, which are periodic in time with the same period as the drive:
\begin{equation}
    | \Psi_\alpha (t) \rangle = e^{-i\epsilon_\alpha t} | \Phi_\alpha (t) \rangle,\quad | \Phi_\alpha (t+T) \rangle=| \Phi_\alpha (t) \rangle .\label{eq:ansatz1}
\end{equation}
Inserting this ansatz in the time-dependent Schrödinger equation, one finds the Floquet equation:
\begin{equation}
    \left[H\left(t\right)-i\partial_{t}\right]|\Phi_{\alpha}\left(t\right)\rangle=\epsilon_{\alpha}|\Phi_{\alpha}\left(t\right)\rangle, \label{eq:FloquetEquation}
\end{equation}
which is an eigenvalue equation for the Floquet states with Floquet operator $K(t)\equiv H(t)-i\partial_t$ and eigenvalues given by the quasienergies.

Typically in Floquet engineering, due to their time-periodicity, one does a Fourier expansion of the Floquet states $|\Phi_{\alpha}\left(t\right)\rangle=\sum_n e^{-i n \omega t}|\Phi_{\alpha,n}\rangle$, and of the Floquet operator, $K(t)$, which allows to write the eigenvalue equation in terms of Fourier components in an extended Hilbert space, also known as Sambe space~\cite{GRIFONI1998}.
The advantage is that the time-dependence effectively disappears, but the price to pay is that the Floquet operator becomes an infinite dimensional matrix.

Fortunately, in the high frequency regime, the matrix approximately is block diagonal, which allows for simple solutions.
However, it is important to stress that the Fourier expansion of the Floquet states is not strictly necessary. It is just specially helpful in the high frequency regime or when the Floquet operator can be block diagonalized~\cite{vogl_flow_2019,thomson_flow_2020}.
\begin{figure}
    \centering
    \includegraphics[width=\columnwidth]{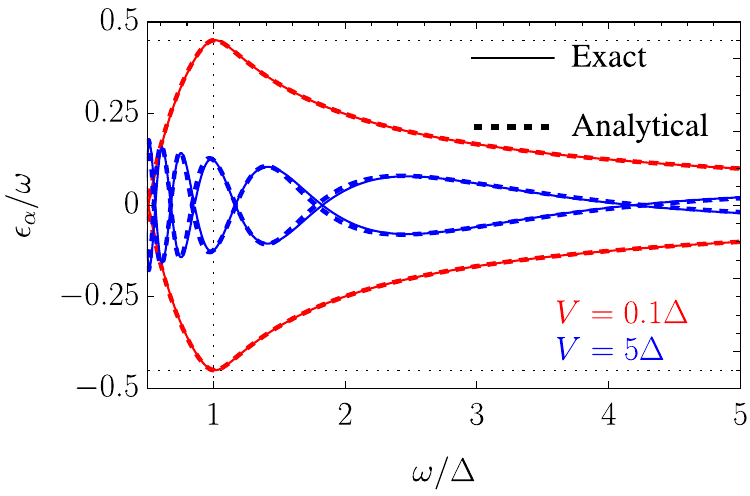}
    \caption{Comparison between the exact (solid) and the analytical quasienergies (dashed), as a function of frequency and for $V=0.1 \Delta$ (red) and $5\Delta$ (blue). The vertical dotted line indicates the value $\omega=\Delta$ and the horizontal dotted line the expected splitting from $\Delta_1$.}
    \label{fig:quasienergies0}
\end{figure}

As we are interested in the role of resonances, we directly work with Eq.~\eqref{eq:FloquetEquation} and solve it for the RWA Hamiltonian in Eq.~\eqref{eq:HRWA1} by applying the transformation $R_2(t)$.
Then, the solution for the Floquet equation in the original frame reads:
\begin{equation}
    | \Phi_{\pm} \left( t \right) \rangle = R_1(t) e^{-i\frac{\omega}{2} \sigma_{z} t} e^{-i\left(\lambda_{\pm}-\epsilon_{\pm}\right)t}|\phi_{\pm}\rangle. \label{eq:FloquetState1}
\end{equation}
Finally, the quasienergies are obtained by requiring that the Floquet state must be $T$-periodic, $|\Phi_{\pm}\left(T\right)\rangle=|\Phi_{\pm}\left(0\right)\rangle$, which imposes:
\begin{equation}
    \epsilon_{\pm}=\frac{\omega}{2}\left(2n+1\right)+\lambda_{\pm},\quad \forall n\in\mathbb{Z}, \label{eq:quasienergies1}
\end{equation}
where we have used that $R_1(t+T)=R_1(t)$ and that $R_2(t+T)=-R_2(t)$.
Therefore, the Floquet states from Eq.~\eqref{eq:FloquetState1} become:
\begin{equation}
    | \Phi_{\pm} \left( t \right) \rangle = R_1(t) e^{-i\frac{\omega}{2} \sigma_{z} t} e^{i\frac{\omega}{2}\left(2n+1\right)t} |\phi_{\pm}\rangle \ . \label{eq:FloquetState2-1}
\end{equation}
Their dependence on the integer $n$ only indicates the redundancy between different Floquet sidebands due to the $2\pi$ phase-periodicity. 
Hence, if one focuses on the zeroth sideband, i.e. in the range $\epsilon_{\alpha} \in [-\frac{\omega}{2},\frac{\omega}{2}]$ of Eq.~\eqref{eq:quasienergies1}, we can finally write the quasienergies and the Floquet states, under the RWA approximation, as:
\begin{align}
    \epsilon_{\pm} &= \pm\frac{\omega}{2}+\lambda_{\mp} \ , \label{eq:QuasienergyRabi} \\
    | \Phi_{\pm} \left( t \right) \rangle &= R_1(t) e^{-i\frac{\omega}{2} (\sigma_{z}\pm 1) t} |\phi_{\pm}\rangle . \label{eq:FloquetState2}
\end{align}
Crucially, notice that in Eq.~\eqref{eq:QuasienergyRabi} the subscript of the quasienergies, $\epsilon_{\pm}$, and of the eigenvalues of the rotating Hamiltonian, $\lambda_\mp$, are reversed in our convention. 
This will be relevant to correctly identify the topological invariants for each band in the examples below.
Also, our convention fixes the definition of $0$- and $\pi$- gaps, being the latter the one that separates the quasienergy in one sideband to the quasienergy of the neighboring one.
\begin{figure*}
    \centering
    \includegraphics[width=0.75\textwidth]{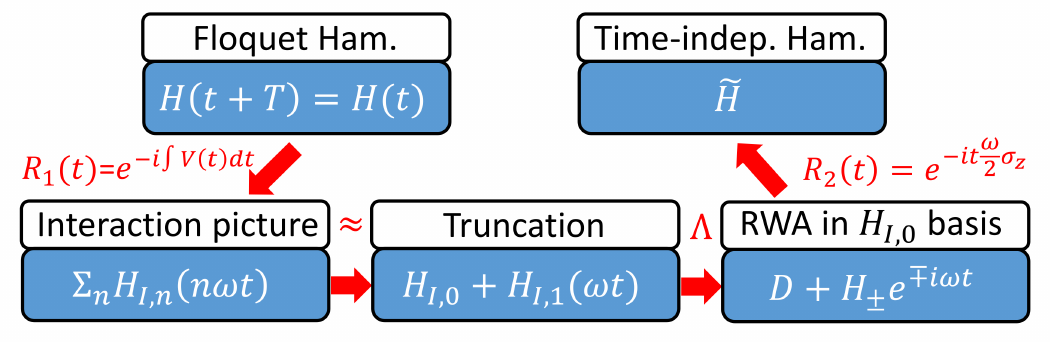}
    \caption{Diagram of the different transformations being used. The original Floquet Hamiltonian is first transformed to the interaction picture to account for strong field effects using $R_1 (t)$. The truncation to the static plus the dominant time-dependent part defines the transformation $\Lambda$, which diagonalizes $H_{0,I}$ and encodes the $0$-gap topology. Finally, a RWA plus a transformation $R_2(t)$ leads to $\tilde{H}$, which encodes the $\pi$-gap topology.}
    \label{fig:Flow1}
\end{figure*}

For illustrative purposes and as a check, let us now particularize to the high frequency regime, $\omega \gg \Delta_1$. There, we can perform a series expansion of the eigenvalues $\lambda_\pm$ and find:
\begin{equation}
    \lambda_\pm = \pm\frac{\omega}{2}\mp \frac{\Delta_0}{2} \pm \frac{\Delta_1^2}{4\omega}+\ldots , \label{eq:TLS-Eigen}
\end{equation}
which simplifies the quasienergies to:
\begin{equation}
    \epsilon_\pm =\pm \frac{\Delta_0}{2} \mp \frac{\Delta_1^2}{4\omega}+\ldots . \label{eq:quasienergies2}
\end{equation}
This result is in agreement with the standard high frequency analysis of the original Hamiltonian in Eq.~\eqref{eq:Rabi1} using a Magnus expansion~\cite{eckardt_high-frequency_2015}, confirming the correctness of our expressions in that regime. 
In addition, our result is more general, as it is valid also at resonance. Hence, we can use this approach  to explore the behavior of the system as the frequency is decreased from the high frequency regime to the resonant one.
To confirm this, we plot in Fig.~\ref{fig:quasienergies0} a comparison between the exact quasienergies and those obtained in Eq.~\eqref{eq:QuasienergyRabi} for the zeroth sideband, for different field amplitudes.
The agreement is excellent not only at the high frequency regime, but also at frequencies below the original resonance, $\omega=\Delta$. Furthermore, the quasienergies are captured for both, the weak coupling (red) and the strong coupling regime (blue).

For a better understanding, let us write the explicit expression for the quasienergies from Eq.~\eqref{eq:QuasienergyRabi}:
\begin{equation}
    \epsilon_\pm = \pm \frac{\omega}{2} \mp \frac{1}{2}\sqrt{\left(\Delta_{0}-\omega\right)^{2}+\Delta_{1}^{2}}
\end{equation}
First, it shows that as a large field amplitude $V$ produces a renormalization of the bands via the Bessel functions, the frequency value at which the resonance happens must be obtained from the condition $\omega=\Delta_0$, which leads to the non-linear equation:
\begin{equation}
    \omega=\Delta \mathcal{J}_0\left( 2V/\omega \right) . \label{eq:freq0}
\end{equation}
The solution to Eq.~\eqref{eq:freq0} provides the value at which the frequency of the drive matches the renormalized splitting of the RWA Hamiltonian.\\
Second, note that due to the contribution $\pm\omega/2$ in Eq.~\eqref{eq:QuasienergyRabi}, the quasienergies near resonance approach each other at the $\pi$-gap~\footnote{The difference between the eigenvalues of the rotated frame Hamiltonian $\lambda_\pm$, which close the gap at zero, and the quasienergies $\epsilon_\pm$, which close the gap at $\pi$, will be important to understand that the topological properties of $\tilde{H}$ are linked to the $\pi$-gap, while it is the transformation $\Lambda$ that captures the $0$-gap topology.}.
However, as they still have a contribution proportional to $\Delta_{1}$, it will lead to an anti-crossing.
This is shown in Fig.~\ref{fig:quasienergies0}, where the weak coupling case (red line) shows a small splitting at the $\pi$-gap. Its value can be obtained by assuming $\omega=\Delta_0$ in Eq.~\eqref{eq:QuasienergyRabi}, to find the splitting $|\Delta_1|=\Delta \mathcal{J}_{n}\left(2V/\omega\right)$ between neighboring sidebands, as shown in Fig.~\ref{fig:quasienergies0} (horizontal dotted lines).\\
In addition, the strong coupling case $V=5\Delta$ is also well-captured, despite of the important contribution from Bessel functions $\mathcal{J}_{n \geq 2}(2V/\omega)$. 
The reason for this is that the large renormalization of the bands by $\mathcal{J}_0(2V/\omega)$ largely reduces the gap size and one must go to even smaller frequencies to reach resonance. Hence, the system remains in the high frequency regime and the eigenvalues are well-approximated by the first term of Eq.~\eqref{eq:quasienergies2}.

In general we find that the analytical solution starts to fail when $\omega$ is of the order of $\Delta_0/2$, which corresponds to the case where the first harmonic, $2\omega$, becomes resonant.
In that situation it is required to use a different effective Hamiltonian with $\mathcal{J}_2(2V/\omega)$ instead of $\mathcal{J}_1(2V/\omega)$.
Also, for the same reason we find that the case $V \sim \Delta$ is generally the most difficult regime to capture analytically. 

As a summary of this introductory section, \textit{we have shown that a set of suitable transformations and approximations allows to understand in analytical terms the physics of a two-band Floquet system, for a wide regime of amplitudes and frequencies}.
The schematic diagram of Fig.~\ref{fig:Flow1} outlines the transformations and approximations required to carry out the complete analysis.

In particular, the transformation $R_1(t)$ to the interaction picture allows to incorporate strong field effects, after a truncation to the regime of interest has been applied. Typically, this only involves to keep the static and the lowest harmonic contribution of the Fourier series expansion, because it will be the one that first becomes resonant.
Next, a transformation $\Lambda$ to the eigenstates basis of the static part allows to identify the rotating terms and eliminate the counter-rotating ones.
Finally, applying the transformation $R_2(t)$ to a rotating frame allows to find an effective static Hamiltonian $\tilde{H}$ that captures the contribution from the resonance to the quasienergies.

Now we move on to spatially extended systems, and generalize the previous approach to the case of bands with a dispersion relation.
\begin{figure*}
    \centering
    \includegraphics[width=\textwidth]{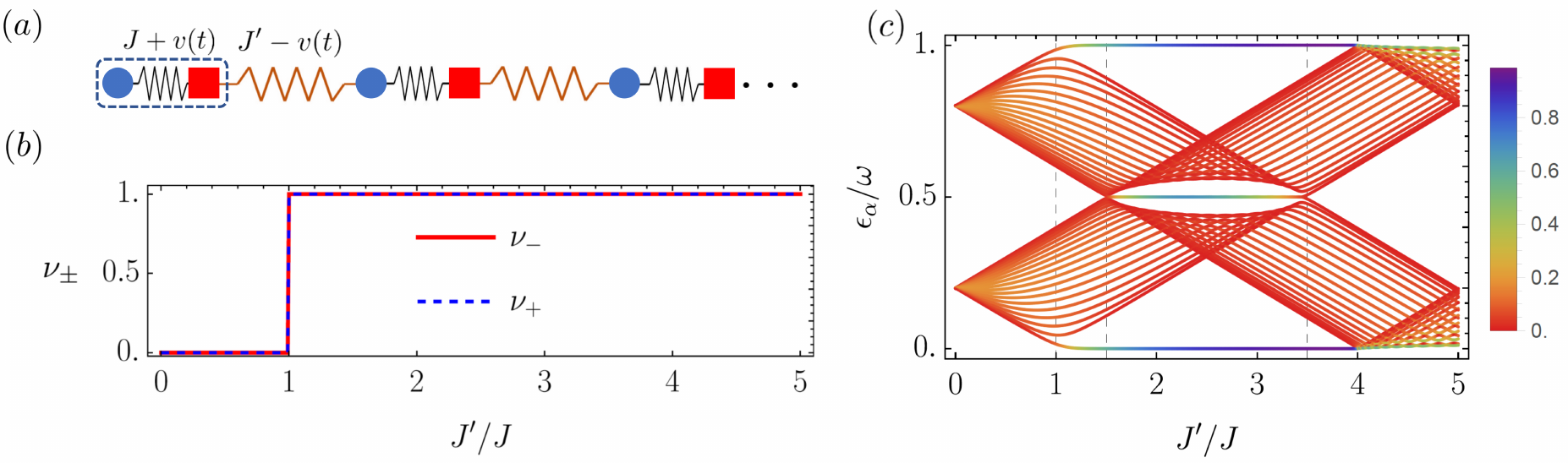}
    \caption{$(a)$ Schematic representation of the periodically driven dimers chain with a chiral symmetric time modulation of the hoppings. The dashed rectangle indicates the unit cell, and the time-modulation linearly enhances/reduces each hopping by changing the separation between sites. $(b)$ Winding number for each band as a function of the dimerization ratio $J^\prime/J$. $(c)$ Quasienergy spectrum centered around the $\pi$-gap for the driven chain with OBC, as a function of $J^\prime/J$. Parameters: $\omega=5J$, $V=0.2J$ and $20$ sites. The color code indicates the localization of the states towards the boundaries.}
    \label{fig:Schematic-SSH}
\end{figure*}
\section{1D Topology: AC driven SSH chain}
Lattice systems in condensed matter describe a wide number of physical models, and when coupled to artificial fields~\cite{Peierls1933,Wannier1962}, it is known that their properties can be externally controlled, giving rise to the field of Floquet engineering~\cite{lindner_floquet_2011,kitagawa_topological_2010,kitagawa_observation_2012,gomez-leon_floquet-bloch_2013,gomez-leon_engineering_2014,delplace_merging_2013,grushin_floquet_2014,rudner_band_2020,NagSynch2019,NagDislocations2021,NagQuadrupolar2022,NagNonHTS2022,NagCascade2022,NagMajorana2023,NagHighOrderTIs2023}.\\
In contrast with the Rabi model previously analyzed, the energy splitting in lattice models generally depends on the quasi-momentum $\mathbf{k}$, due to the dispersive nature of the bands. This is important because, for a fixed frequency, now there will simultaneously be states that are resonant and off-resonant with the drive~\footnote{Single-band models could also display resonant transitions, but only if the driving field is not spatially homogeneous and can couple states with different momentum.}.

In lattice systems, one-dimensional, two-band models represent some of the simplest cases where non-trivial topology can be present, and it is known that when driven out of equilibrium, transitions between the valence and the conduction band can affect the topology.
In particular, a two-band model in one dimension where the driving field can introduce topological changes is found in the Su-Schrieffer-Hegger (SSH) chain~\cite{Su1979}. When it is periodically driven in time~\cite{gomez-leon_floquet-bloch_2013,perez-gonzalez_simulation_2019,Perez-Gonzalez2019-2}, one finds that the quasienergies can exhibit topological edge states in both gaps. Furthermore, if the hopping amplitudes are periodically modulated in a chirally symmetric way, it is possible to find regions of the phase diagram where $0$-gap and $\pi$-gap edge states can coexist~\cite{dal_lago_floquet_2015,balabanov_robustness_2017,cardano_detection_2017}. In that case, the Zak phase of the Floquet bands or winding number (if expressed in units of $\pi$), is trivial. However, it is known that one can define a new pair of topological invariants adapted to Floquet phases, and topologically characterize this anomalous phase~\cite{asboth_chiral_2014}, predicting the existence of edge states. We will refer to this situation, which does not have a static analog, as the anomalous topological phase of this 1D Floquet Hamiltonian.

Next, we discuss the topology in different frequency regimes.
\subsection{High frequency topology}
The Hamiltonian for the periodically driven SSH chain can be written as:
\begin{equation}
    H\left(t\right) = \sum_{j}\left(J_{1}\left(t\right)b_{j}^{\dagger}a_{j}+J_{2}\left(t\right)a_{j+1}^{\dagger}b_{j}+\text{h.c.}\right) , \label{eq:SSH1}
\end{equation}
with $J_{1}\left(t\right)\equiv J+v\left(t\right)$, $J_{2}\left(t\right)\equiv J^{\prime}-v\left(t\right)$ and the modulation of the hopping $v\left(t\right)=2V\cos\left(\omega t\right)$, being $V$ the modulation amplitude and $\omega$ its frequency.

This time modulation is different to the one typically assumed for light-driven Hamiltonians~\cite{gomez-leon_floquet-bloch_2013,Perez-Gonzalez2019-2}. The system can be thought as a dimerized lattice with sites linearly coupled by springs, and with the hopping of particles directly proportional to the distance between them. When the sites in one sublattice are linearly shifted from their equilibrium position, one hopping is enhanced the same amount as the other one is reduced. This modulation is harmonically changed with frequency $\omega$, related with the stiffness of the springs (see Fig.~\ref{fig:Schematic-SSH}, panel $(a)$, for an schematic)~\footnote{The springs could be lattice vibrations in some realizations, if their frequency is in the correct range of energies.}.

To analyze the system in simple terms, we first consider the case of periodic boundary conditions (PBC). The system becomes translationally invariant and we can Fourier transform to momentum space. 
In the basis $\left(a_{k},b_{k}\right)^{T}$, the time-dependent Hamiltonian matrix reads:
\begin{equation}
    H_k(t) = J_{1}\left(t\right)\sigma_{x}+J_{2}\left(t\right)\left(\sigma_{+}e^{-ik}+\sigma_{-}e^{ik}\right), \label{eq:Hdimers1}
\end{equation}
with $\sigma_\pm = \frac{1}{2}(\sigma_x \pm i\sigma_y)$. 
As we are mostly interested in the topological changes introduced by the resonance, we consider weak field amplitudes $V<\omega$, such that the transformation to the interaction picture $R_1 (t)$ becomes unnecessary (see Fig.~\ref{fig:Flow1} for a reminder of the different transformations involved). This is not required, but it greatly simplifies the analysis and enlightens the role of resonances.

The high frequency topology is easily captured from the effective stroboscopic Hamiltonian, which is given by the average Hamiltonian $H_0(k) \equiv \int_{0}^{T}H_{k}\left(t\right)dt/T$~\cite{eckardt_high-frequency_2015}:
\begin{equation}
    H_0(k) = \left(\begin{array}{cc}
0 & J+J^{\prime}e^{-ik}\\
J+J^{\prime}e^{ik} & 0
\end{array}\right) \label{eq:Hdimers2} .
\end{equation}
That is, for our case with weak field amplitudes, it coincides with the undriven SSH chain Hamiltonian.


Due to chiral symmetry $\{ \sigma_z, H_0(k) \}=0$, the Zak phase is quantized and can be calculated in terms of the parallel transport of eigenstates:
\begin{equation}
    \mathcal{Z}_\pm = \int_{-\pi}^\pi \langle v_\pm (k) | i\partial_k |v_\pm (k) \rangle dk , \label{eq:Zak0}
\end{equation}
being $E_\pm(k)$ the eigenvalues of $H_0(k)$ and $|v_\pm (k) \rangle$ the eigenstates, such that $H_0(k)|v_\pm (k)\rangle = E_\pm(k) |v_\pm (k) \rangle$.

At this step, it will be useful to rewrite Eq.~\eqref{eq:Zak0} in terms of the matrix that diagonalizes the average Hamiltonian $\Lambda(k)^\dagger H_0(k) \Lambda(k) = D(k)$, with $D(k)$ a diagonal matrix with entries $E_\pm(k)$.
The Zak phase from Eq.~\eqref{eq:Zak0} becomes:
\begin{equation}
    \mathcal{Z}_\pm = \int_{-\pi}^\pi \langle \pm | i\partial_k \log \Lambda(k) |\pm \rangle dk , \label{eq:Zak1}
\end{equation}
with $| \pm \rangle$ the unit vectors $(1,0)$ and $(0,1)$, identifying the conduction and valence band.
Its calculation is straightforward using a contour in the complex plane, and due to its relation with the winding number of the system $\nu_{\pm} = \mathcal{Z}_\pm/\pi$, we can directly write $\nu_{\pm} = \Theta\left(J^{\prime}/J-1\right)$.
This result is shown in Fig.~\ref{fig:Schematic-SSH}, panel $(b)$, and as expected, confirms that the high frequency topology for weak field is identical to that of the undriven SSH chain. There is a topological phase transition at $J=J^\prime$, and in the absence of renormalization effects due to strong field effects, the periodic field does not affect the topology.\\
Importantly, note that the winding number for the two bands is identical and that all the topological properties in Eq.~\eqref{eq:Zak1} are now encoded in the $k$-dependence of $\Lambda(k)$, because the unit vectors $|\pm\rangle$ only select the band.

To confirm our results and check their regime of validity, in Fig.~\ref{fig:Schematic-SSH}, panel $(c)$, we plot the exact quasi-energies of $H_k(t)$, centered at the $\pi$-gap, for the case of Open Boundary Conditions (OBC).
It shows that for small $J^\prime/J$, the bandwidth is smaller than the frequency $2(J+J^\prime)<\omega$, and the system remains in the high-frequency regime.
For that reason topological edge state appear in the $0$-gap at $J^\prime=J$ (vertical dashed line), as predicted by the winding number of the stroboscopic Hamiltonian $\nu_\pm$.

For larger values of $J^\prime/J$, the bandwidth continues to increase until $J^\prime=1.5J$, where it equals the drive frequency and the system abandons the high-frequency regime, making our predictions invalid. Entering in this resonant regime the $\pi$-gap closes and an additional localized edge state in the $\pi$-gap emerges, which is not captured by the high frequency stroboscopic Hamiltonian $H_0(k)$. Interestingly, when the edge states in both gaps are simultaneously present, the exact numerical calculation of the winding number from the Floquet states results in a zero value.

In the next section, we obtain an effective Hamiltonian that incorporates the physics of the resonance, discuss the appearance of the $\pi$-gap edge states and show their topological origin.
\subsection{Derivation of the RWA Hamiltonian}
\begin{figure*}
    \centering
    \includegraphics[width=0.8\textwidth]{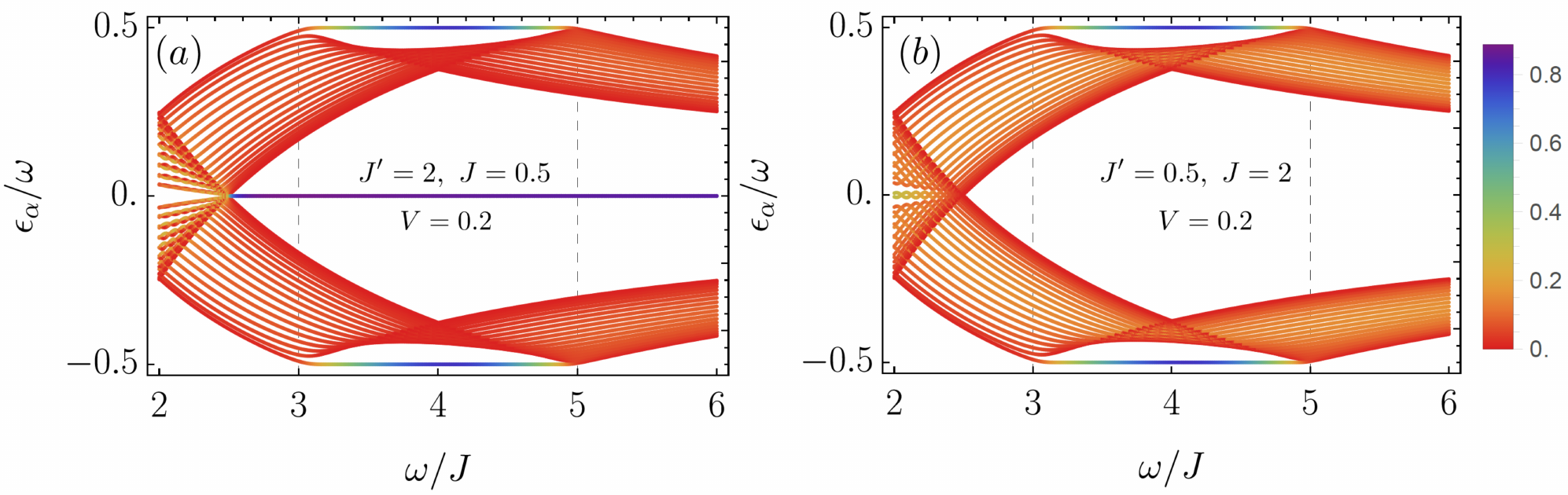}
    \caption{Quasienergies for OBC as a function of frequency and centered at the $0$-gap. The color code indicates the localization towards the edges. $(a)$ Spectrum for the case of a topological phase at high frequency. $(b)$ Spectrum for the case of a trivial phase at high frequency. The critical points are independent of the high frequency topology and only depend on the resonances at $\omega=2|J \pm J^\prime|$.}
    \label{fig:ResonanceDimers1}
\end{figure*}
As was discussed in the previous section, precisely at resonance the system develops additional edge states in the $\pi$-gap. This indicates that resonances play a crucial role in the anomalous topology of Floquet phases. 
To further understand this process, we will derive a RWA Hamiltonian for the driven SSH chain, that captures the main mechanism causing the topological phase transition.
This will be done following the diagram sketched in Fig.~\ref{fig:Flow1}, but skipping the transformation to the interaction picture $R_1(t)$ and the posterior truncation. This is possible because, for this model, we are assuming a weak coupling regime, to focus on the role of resonances.

We consider the Hamiltonian with PBC and separate the driven and static parts of the Hamiltonian:
\begin{equation}
    H_{k}\left(t\right)=H_{0}\left(k\right)+V\left(k,t\right) ,
\end{equation}
with
\begin{equation}
    V(k,t) \equiv v(t) \left[1-\cos\left(k\right)\right]\sigma_{x}-v(t) \sin\left(k\right)\sigma_{y} .
\end{equation}
As a resonance is typically defined in terms of a transition between the conduction and the valence band, we first diagonalize the static part of the Hamiltonian:
\begin{equation}
    \Lambda^{\dagger}(k)H_{0}(k)\Lambda(k)=D(k) , \label{eq:Diagonalization1}
\end{equation}
being $D(k)$ a diagonal matrix with the energies $E_\pm(k)=\pm\sqrt{J^{2}+J^{\prime2}+2JJ^{\prime}\cos\left(k\right)}$ as entries and $\Lambda(k)$ a unitary transformation matrix with its columns given by the eigenvectors of $H_{0}(k)$.
Notice that $\Lambda(k)$ encodes all the information about the topology of the effective stroboscopic SSH chain Hamiltonian in its $k$-dependence, and that the energies $E_{\pm}(k)$ provide a definition for the bands where the resonances will take place.

We rewrite $H_k(t)$ in this eigenstates basis
\begin{equation}
    \hat{H}_k(t) \equiv \Lambda^{\dagger}(k)H_k(t)\Lambda(k)=D(k)+\hat{V}(k,t) , \label{eq:Diagonalization2}
\end{equation}
with $\hat{V}(k,t)=\Lambda^{\dagger}(k)V(k,t)\Lambda(k)$ now being a complicated matrix with all its entries different from zero. 
To simplify its form, one needs to remember how different physical processes are related with the various entries. For example, the diagonal part does not couple different bands and corresponds to a time-modulation, independent for each band, that only modifies the phase acquired over time by each eigenstate, in a non-linear way.\\
Off-diagonal elements are more relevant because they couple different bands. They can be separated in two types: the ones that can produce resonances $\hat{V}_\text{R}(k,t)$, and the ones that cannot $\hat{V}_\text{NR}(k,t)$. For a two-band model they are easy to identify, as the resonant ones coincide with the rotating terms of the Rabi model:
\begin{equation}
    \hat{V}_\text{R}(k,t) = \hat{V}(k)^{2,1} \sigma_{-} e^{i\omega t} + \hat{V}(k)^{1,2} \sigma_{+} e^{-i\omega t} , \label{eq:RWA1}
\end{equation}
with $j$ and $l$ in $\hat{V}(k)^{j,l}$ indicating the row and column of the time-independent part of the matrix $\hat{V}(k,t)$, respectively.
In more complex cases with a larger number of bands the resonant terms can be systematically identified with the secular terms of the first order correction in time-dependent perturbation theory~\cite{gomez-leon_designing_2020,Gomez-Leon-2022-Multiqudit}:
\begin{equation}
    \int e^{iD(k)t}\hat{V}\left(k,t\right)e^{-iD(k)t}dt \ .
\end{equation}
All the remaining time-dependent, off-diagonal terms of $\hat{V}(k,t)$ can be identified with counter-rotating terms. Their contribution is small near resonances and their effect can be incorporated by means of perturbation theory later on. For this reason, to lowest order we will keep the RWA terms only and define our approximate Hamiltonian for the SSH chain as:
\begin{equation}
    H_{\text{RWA}}(k,t)=D(k)+\hat{V}_R(k,t)\ .
\end{equation}
In this particular case it is easy to write the explicit form of the RWA terms (they fulfill $\hat{V}(k)^{1,2}=[\hat{V}(k)^{2,1}]^*$) as:
\begin{equation}
    \Gamma(k) \equiv \hat{V}(k)^{1,2} = \frac{i (J+J^{\prime}) V \sin\left(k\right)}{\sqrt{J^{2}+J^{\prime2}+2JJ^{\prime}\cos\left(k\right)}} , \label{eq:off-diagonal1}
\end{equation}
which vanish for $V\to0$, as expected, but also at $k=n\pi$ with $n\in\mathbb{Z}$.
\begin{figure*}
    \centering
    \includegraphics[width=0.8\textwidth]{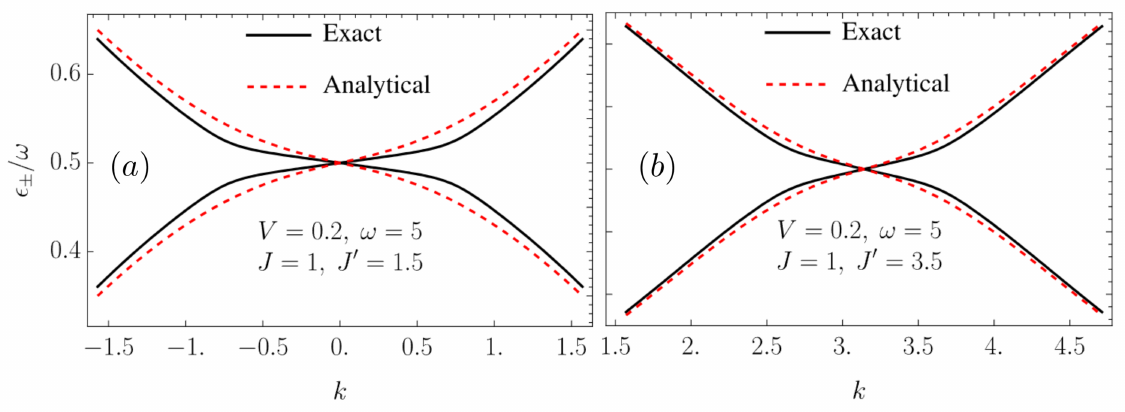}
    \caption{Comparison between the exact and the analytical quasienergies for the system with PBC at resonance for $J^\prime=1.5$ $(a)$ and $3.5$ $(b)$. The $\pi$-gap closes at high symmetry points $k=0$ and $\pi$. Notice the scale of the plot where differences are enhanced.}
    \label{fig:ResonanceDimers2}
\end{figure*}

We can now exactly solve the time-dependent Schr\"{o}dinger equation for the $H_{\text{RWA}}(k,t)$ in a similar way as for the Rabi model, by just applying a time-dependent transformation $R_2(t)=e^{-i\frac{\omega}{2}t\sigma_{z}}$. This results in the following \emph{time-independent} Hamiltonian in the rotating frame:
\begin{equation}
    \tilde{H}\left(k\right)	= \left(\begin{array}{cc}
E_{+}\left(k\right)-\frac{\omega}{2} & \Gamma\left(k\right)\\
\Gamma\left(k\right)^{*} & \frac{\omega}{2}-E_{+}\left(k\right)
\end{array}\right) . \label{eq:RotatingFrameH1}
\end{equation}
Notice its similarity with Eq.~\eqref{eq:HRWA1-1}, where the frequency also changes the gap of the unperturbed bands.

As was discussed for the Rabi model, the effective Hamiltonian $\tilde{H}$ describes the physics in the $\pi$-gap, while the topology of the $0$-gap is encoded in $\Lambda(k)$. In particular, when we calculate the eigenvalues
\begin{equation}
    \tilde{E}_{\pm}(k) = \pm \sqrt{\left( E_{+}\left(k\right)-\frac{\omega}{2} \right)^2 + |\Gamma(k)|^2}, \label{eq:eigenvalues2}
\end{equation}
one can see that, as in the Rabi model, for the resonance condition
\begin{equation}
    \omega=2E_+(k)=2\sqrt{J^{2}+J^{\prime2}+2JJ^{\prime}\cos\left(k\right)} , \label{eq:Resonance1}
\end{equation}
the rotating terms produce anti-crossings of size proportional to $|\Gamma(k)|$ between neighboring quasienergies.
However, in contrast with the Rabi model, at particular points of the FBZ given by $k=n\pi$, the eigenvalues become exactly degenerate and the $\pi$-gap closes.\\
Note that the existence of an exact degeneracy is a consequence of the symmetries of the interplay of both, the bands and the driving field.
Then, as exact crossings are important because they can produce topological phase transitions, this is what allows the coexistence of $0$-gap and $\pi$-gap edge states in the driven SSH chain.
For this reason, different driving protocols might fail to produce a robust anomalous phase in the driven SSH chain.

To demonstrate this explicitly, let us first find the condition for a gap closure as a function of frequency. As it must happen at the particular points $k=n\pi$, because only there $\Gamma(k)$ in Eq.~\eqref{eq:eigenvalues2} vanishes, the resonance condition in the first term simplifies to $\omega=2E_{+}(n\pi)$.
In particular, for each case we obtain:
\begin{equation}
    \omega=\begin{cases}
2\left(J+J^{\prime}\right) & \text{for }k=0\\
2\left|J-J^{\prime}\right| & \text{for }\left|k\right|=\pi \label{eq:Resonance-cond}
\end{cases}
\end{equation}
These two values perfectly predict the closures of the $\pi$-gap in Fig.~\ref{fig:Schematic-SSH}, first at $J^\prime=1.5J$ and then at $J^\prime=3.5J$ (vertical dashed lines), linked with the appearance and disappearance of the edge states in that gap.
Importantly, their position does not change for moderate values of $V$, confirming that the resonance mechanism is what controls the topological phase transition and not strong field renormalization effects.

Equation~\eqref{eq:Resonance-cond} also indicates that the gap closure does not depend on the dimerization phase. Hence, the appearance/disappearance of edge states in the $\pi$-gap must be independent of the topology of the high frequency stroboscopic Hamiltonian.
We confirm that this is indeed the case, as shown in Fig.~\ref{fig:ResonanceDimers1}, panels $(a)$ and $(b)$. One can see that as the frequency is reduced, the gap closures are perfectly predicted by Eq.~\eqref{eq:Resonance-cond} for both cases, independently of the original dimerization phase, with the corresponding appearance/disappearance of edge states.
This feature indicates that the presence of edge states in each gap can be independently studied, in agreement with the definition of invariants in Floquet systems for each gap~\cite{asboth_chiral_2014,cheng_observation_2019}.

In Fig.~\ref{fig:ResonanceDimers2}, panels $(a)$ and $(b)$, we compare a zoom of the exact quasienergy spectrum for PBC, with the analytical one from Eq.~\eqref{eq:eigenvalues2}. One can see their good agreement, and importantly, confirm that the rotating terms dominate near resonance, because the ratio between counter-rotating and rotating terms, or equivalently, the disagreement with the exact result gets reduced as $k$ approaches $k\to n\pi$.
\subsection{Topology induced by resonances}
Finally, as we have analytical expressions for the eigenvalues and eigenstates of $\tilde{H}(k)$, that perfectly predict the appearance of edge states in the $0$- and $\pi$-gaps, let us analyze the Zak phase of the corresponding Floquet states.
In the original frame of reference they are:
\begin{equation}
    |\Phi_{\pm}\left(k,t\right)\rangle = \Lambda \left(k\right)e^{-i\frac{\omega}{2}t (\sigma_{z}\mp 1)}|\phi_{\pm}\left(k\right)\rangle, \label{eq:FloquetStateDimer}
\end{equation}
where $|\phi_\pm(k)\rangle$ are the eigenstates of $\tilde{H}(k)$, the exponential $e^{-i\frac{\omega}{2}(\sigma_z \mp 1)t}$ undoes the transformation to a co-rotating frame, restricting the Floquet states to the first sideband, and $\Lambda(k)$ encodes the $k$-dependence of the conduction and valence bands.

Note that the only difference with the Floquet states in the Rabi model is the additional $k$-dependence in both, the transformation $\Lambda(k)$ and the eigenstates $|\phi_\pm(k)\rangle$.
Importantly, as the Floquet states are periodic in both, $k$ and $t$, their phases can be compared after a periodic cycle and they can be used to characterize the full topological properties~\cite{Aharonov1987}.
Then, as the $\pi$-gap closure is captured by the rotating frame Hamiltonian $\tilde{H}(k)$, we expect that $| \phi_\pm (k) \rangle$ encodes the $\pi$-gap topology. In contrast, the $0$-gap topology is related with the off-resonant physics, and then, it must be related with the transformation $\Lambda(k)$.

\begin{figure}
    \centering
    \includegraphics[width=\columnwidth]{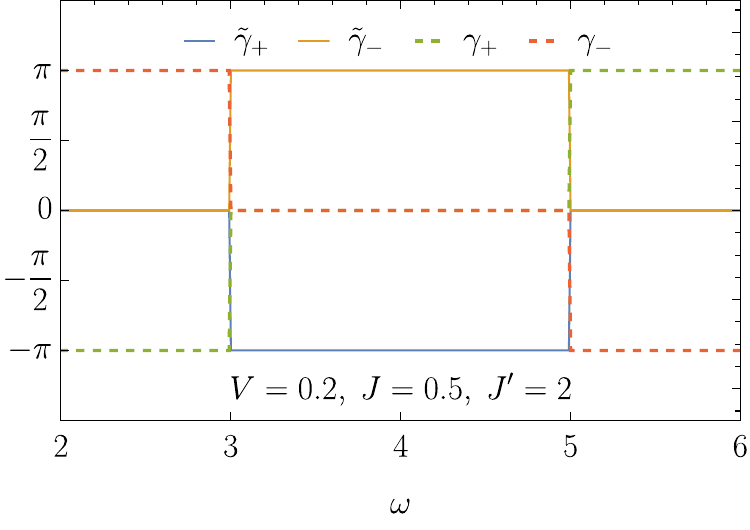}
    \caption{Dashed lines show the Zak phase $\gamma_\pm$ of each Floquet state, as a function of the frequency (dashed), for arbitrary time $t$. The solid lines correspond to the contribution from the rotating frame Hamiltonian $\tilde{\gamma}_\pm$. Parameters: $V=0.2$, $J=0.5$ and $J^\prime=2$.}
    \label{fig:ZakDimers}
\end{figure}

To demonstrate this explicitly, we calculate the Zak phase:
\begin{equation}
    \gamma_{\pm}(t) = \int_{-\pi}^{\pi} \langle\Phi_{\pm}\left(k,t\right)|i\partial_{k}|\Phi_{\pm}\left(k,t\right)\rangle dk , \label{eq:Zak}
\end{equation}
and if we use Eq.~\eqref{eq:FloquetStateDimer}, we can separate the result in two contributions:
\begin{equation}
    \gamma_{\pm}(t) = \tilde{\gamma}_\pm + \bar{\gamma}_\pm (t) , \label{eq:ZakEq}
\end{equation}
where we have defined the \emph{time-independent} Zak phase in the rotating frame
\begin{equation}
    \tilde{\gamma}_\pm \equiv \int_{-\pi}^{\pi} \langle\phi_{\pm}\left(k\right)|i\partial_{k} \phi_{\pm}\left(k\right)\rangle dk , \label{eq:ZakRotating}
\end{equation}
and
\begin{equation}
    \bar{\gamma}_\pm (t) \equiv \int_{-\pi}^{\pi}\langle\phi_{\pm}(k,t)| i\partial_{k}\log\Lambda (k) |\phi_{\pm}(k,t)\rangle dk , \label{eq:ZakFloq}
\end{equation}
which is similar to Zak phase from the stroboscopic Hamiltonian in Eq.~\eqref{eq:Zak1}, now evaluated between Floquet states $|\phi_{\pm} (k,t) \rangle \equiv e^{-i\frac{\omega}{2}t\left(\sigma_{z}\mp1\right)}|\phi_{\pm}(k)\rangle$ instead.\\
In Fig.~\ref{fig:ZakDimers} we plot, in dashed lines, the value of the Zak phase from Eq.~\eqref{eq:Zak} as a function of frequency, for an arbitrary time $t$~\footnote{Although the second contribution to Eq.~\eqref{eq:ZakEq} is time dependent, its evaluation turns out to be time independent.}. The solid lines correspond to the value of the Zak phase in the rotating frame Hamiltonian, Eq.~\eqref{eq:ZakRotating}.
The calculations have been performed by discretizing the momentum and computing the total angle rotated by the overlap function $\langle \Phi_\pm (k,t)| \Phi_\pm (k + \delta k,t) \rangle$, as the FBZ is swept. Also, note that the parameters have been chosen such that the undriven SSH chain is in the topological phase.

Interestingly, Fig.~\ref{fig:ZakDimers} shows that \textit{the Zak phase in the rotating frame, $\tilde{\gamma}_\pm$, is quantized, vanishes in the high frequency regime, and only changes its value when the $\pi$-gap closes its gap due to a resonance}. 
Hence, it perfectly predicts the existence of edge states in the $\pi$-gap and endows them with a topological origin.\\
In contrast, the total Zak phase $\gamma_\pm$ in Fig.~\ref{fig:ZakDimers} only changes its value when $\tilde{\gamma}_\pm$ does, and it goes to zero when the system is in the anomalous phase (i.e., when edge states are simultaneously present in both gaps).
This shows that, although neither $\tilde{\gamma}_\pm$ nor $\bar{\gamma}_\pm$ vanish in the anomalous phase, their combination in Eq.~\eqref{eq:ZakEq} can be incorrectly interpreted as a topologically trivial phase.\\
Importantly, note that the high frequency behavior is characterized by $\bar{\gamma}_\pm$ (because $\tilde{\gamma}_\pm$ vanishes). This confirms that the topology for large frequencies is encoded in $\Lambda(k)$. However, its evaluation between Floquet states in Eq.~\eqref{eq:ZakFloq} results in a different sign for each band. This nicely combines with the opposite sign of $\tilde{\gamma}_\pm$ in the resonant regime.
Also, we have checked that the value of $\bar{\gamma}_\pm$ depends only on the ratio $J^\prime/J$, as expected from $\Lambda(k)$.\\
To confirm these findings, we have calculated the Zak phase for the exact Floquet states from the full, periodically driven SSH Hamiltonian and found perfect agreement with Fig.~\ref{fig:ZakDimers} for the frequency range $\omega\in\left[ 2.5,\infty \right)$. Below this range, additional crossings at both gaps affect the Zak phase.
To capture the topological changes induced by them, it is necessary to include the transformation to the interaction picture $R_1(t)$ and higher harmonics of the driving field.\\

%

In summary, our results explain why the Zak phase in a resonant situation results in a topologically trivial band, but the system displays topological edge states in both gaps.
We have shown that the topological invariant actually corresponds to the sum of two contributions:
\begin{itemize}
    \item An invariant of the rotating frame Hamiltonian, which captures the resonance mechanism and characterizes the $\pi$-gap topology.
    \item An invariant of the original lab frame Hamiltonian, which captures the off-resonant physics and characterizes the $0$-gap topology.
\end{itemize}
Notice that this identification between frames of reference and gaps avoids the use of logarithm branch cuts, typically present in the study of Floquet systems. 
In our case, we have access to a rotating frame invariant which in combination with the invariant from the Floquet bands allows us to fully determine the topology of the system, in agreement with similar results for the Kitaev chain~\cite{Benito2014}.
It is interesting that this topological characterization does not require one to explicitly make use of the micromotion, although it shows its crucial role, because it is what separates the different frames of reference. This simplifies the mathematical analysis, as the base manifold is just spanned by the FBZ.

A natural question to ask is if these results are valid only for the present case of a 1D topological system with chiral symmetry.
In the following section, we discuss two-dimensional systems and show that a similar analysis can be performed where the general conclusions drawn from the 1D case still hold.
\section{2D Topology: AC driven \texorpdfstring{$\pi$}{TEXT}-flux lattice}
\begin{figure*}
    \centering
    \includegraphics[width=0.86\textwidth]{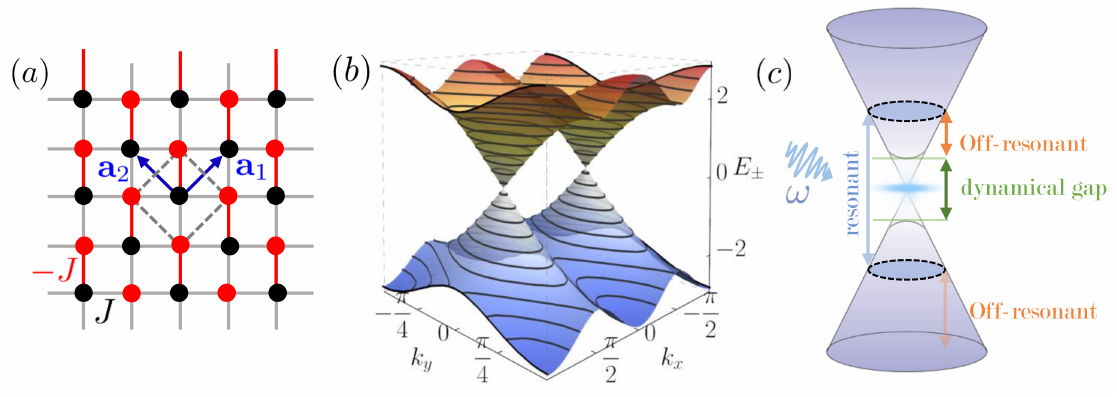}
    \caption{$(a)$ Schematic of a $\pi$-flux lattice with alternating hopping $\pm J$ along the vertical axis, unit vectors $\mathbf{a}_{1,2}$ and lattice constant $a=1$. $(b)$ Band structure of the undriven system and its pair of Dirac cones at $\mathbf{k}=\pi(\pm 1, 0)/2$. $(c)$ Schematic description of a driven Dirac cone and the relevant energy scales.}
    \label{fig:Schematic-Square}
\end{figure*}
Anomalous Floquet topological phases are typically associated with two-dimensional systems, where the Chern number of the quasienergy bands vanishes, but the presence of chiral edge states in the two unequivalent gaps indicates some underlying topology~\cite{rudner_band_2020}.
In previous works it was shown that their topological analysis can be performed in terms of a three-dimensional winding number $W[U_\epsilon]$~\cite{Rudner_2013}, which is evaluated over the $(\mathbf{k},t)$ manifold. Therefore, a non-vanishing value indicates that the micro-motion is a fundamental ingredient for their characterization~\cite{nathan_topological_2015}, because during the dynamics within a period the bands can exhibit topological phase transitions.

In this section, we generalize our previous approach to two dimensions and show that resonances still control the topological phase transitions in the $\pi$-gap.
We demonstrate that the Chern number of the Floquet bands can be written as the sum of two time-independent invariants in different frames related by the presence of a micromotion.

Concretely, we consider a square lattice with alternating hopping along the vertical axis (see Fig.~\ref{fig:Schematic-Square}, panel $(a)$, for a schematic description).
This bipartite lattice can be interpreted as a square lattice with a magnetic flux of half a quanta along the vertical axis~\cite{Zhang2020,Rodriguez-Vega-2019}.
The band structure has two unequivalent Dirac cones and, in the absence of driving, its topology is similar to the one of a honeycomb lattice, with two topological Dirac points protected by symmetry~\cite{BernevigBook}. For this reason, when OBC are considered, the spectrum displays zero-energy modes localized at the edges, whose appearance depends on the particular boundary under consideration~\cite{delplace_zak_2011}.

An advantage of the $\pi$-flux with respect to other lattices, such as the honeycomb, is that when coupled to an AC field via the Peierls substitution, its Floquet operator is quite direct to write. 
On top of this, the $\pi$-flux lattice has been less studied in the literature, which is why we think this model is interesting to exemplify our findings and perform a full analysis of its Floquet physics.
\subsection{High frequency topology}
When a high frequency AC field is applied to a Dirac semi-metal such as this one, it is possible to open a Haldane gap and generate chiral edge states in the $0$-gap~\cite{kitagawa_topological_2010,lindner_floquet_2011,delplace_merging_2013,Diaz2019}.
As the frequency is reduced, the gap between Floquet sidebands can be closed and additional edge states in the $\pi$-gap might appear, which can be propagating or counter-propagating~\cite{gomez-leon_engineering_2014}.
Importantly, the Chern number of the Floquet bands changes as the frequency is reduced, and typically its value does not coincide with the number of edge states in the sample, invalidating the bulk-to-edge correspondence of static topological insulators~\cite{bessho_nielsen-ninomiya_2021}. This feature indicates that anomalous topological phases are possible. 
The relation between bulk invariants and edge states is only recovered when a generalization of the standard classification includes the role of micromotion~\cite{Rudner_2013,nathan_topological_2015,roy_periodic_2017}.

In particular for our case, the Hamiltonian for the undriven lattice is given by:
\begin{equation}
    H_\mathbf{k}=2J\left[ \cos(k_x) \sigma_x + \sin(k_y) \sigma_y \right] , \label{eq:Undriven-Square-H}
\end{equation}
and its energies simply are $\pm2J\sqrt{\cos^{2}\left(k_{x}\right)+\sin^{2}\left(k_{y}\right)}$, with the two unequivalent Dirac cones located at $\mathbf{k} = (\pm \pi/2,0)$, as shown in Fig.~\ref{fig:Schematic-Square}, panel $(b)$.
The Peierls substitution, $\mathbf{k}\to \mathbf{k}+\mathbf{A}\left(t\right)$, produces the periodically driven model, and in particular, if we assume a simple time-dependence of the form: $\mathbf{A}(t)=(A_{x}\sin\left(\omega t\right),A_{y}\sin\left(\omega t+\varphi\right)$, it is possible to express the Hamiltonian in terms of its Fourier components:
\begin{equation}
    H_{\mathbf{k}} (t)	= \sum_{n}H^{\left(n\right)}_\mathbf{k} e^{i n\omega t},
\end{equation}
For that, one just needs to use the Jacobi-Anger identity, which introduces the familiar Bessel functions, and results in:
\begin{align}
    H_{\mathbf{k}}^{\left(n\right)} =& \left[e^{ik_{x}}\mathcal{J}_{n}\left(A_{x}\right) + e^{-ik_{x}}\mathcal{J}_{-n}\left(A_{x}\right)\right]\sigma_{x} \nonumber \\
&-ie^{in\varphi}\left[e^{ik_{y}}\mathcal{J}_{n}\left(A_{y}\right) - e^{-ik_{y}}\mathcal{J}_{-n}\left(A_{y}\right)\right]\sigma_{y},
\end{align}
with $\mathcal{J}_{n}\left(z\right)$ the $n$-th Bessel function of the first kind.
Notice that the phase difference between components, $\varphi$, or polarization, appears only for components with $n\neq 0$. This means that some dynamics is required to define the polarization, as one would expect.

\begin{figure*}
     \centering
     \includegraphics[width=1\textwidth]{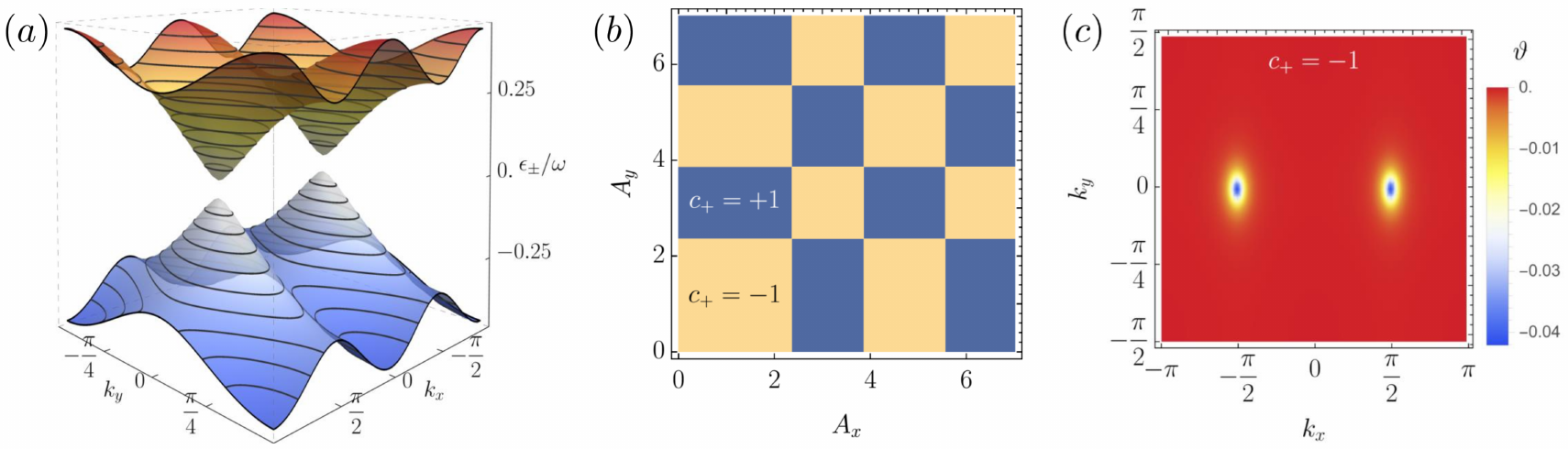}
     \caption{$(a)$ Quasienergies at high frequency for $A_{x} = A_{y} = 0.5$. The mass term opens a topological $0$-gap which leads to a non-zero Chern number and to the appearance of edge states. $(b)$ Topological phase diagram as a function of the field amplitude along each axis. $(c)$ Berry flux distribution for the upper quasienergy band highly localized around the Dirac points. Parameters: $\omega=6J$ and $\varphi = \pi/2$.}
     \label{fig:Square-Chern}
\end{figure*}

As for the SSH chain in the previous section, a first analysis of the topology can be easily carried-out using a Magnus expansion, which is valid in the high frequency regime.
Concretely, it is known that in order to open the Haldane gap, on top of the average Hamiltonian, frequency corrections to first order must be considered~\cite{gomez-leon_engineering_2014,grushin_floquet_2014,Sato2019}.
In that case the stroboscopic Hamiltonian can be approximated by:
\begin{equation}
    \bar{H}_{\mathbf{k}}\approx H_{\mathbf{k}}^{\left(0\right)} + \sum_{n>0}h_{n}\left(\mathbf{k},\varphi\right)\sigma_{z} , \label{eq:Square-HF1}
\end{equation}
with the mass term components $h_{n}\left(\mathbf{k},\varphi\right) \sigma_z = [H_{\mathbf{k}}^{\left(n\right)} , H_{\mathbf{k}}^{\left(-n\right)}] / n \omega$, and in particular,
\begin{equation}
    h_{n}\left(\mathbf{k},\varphi\right) \equiv -\frac{16J_{x,n}J_{y,n}}{n\omega}\sin\left(k_{x}\right)\cos\left(k_{y}\right)\sin\left(n\varphi\right)
\end{equation}
for $n$ an odd integer, and
\begin{equation}
    h_{n}\left(\mathbf{k},\varphi\right) \equiv \frac{16J_{x,n}J_{y,n}}{n\omega}\cos\left(k_{x}\right)\sin\left(k_{y}\right)\sin\left(n\varphi\right)
\end{equation}
for $n$ an even integer. We have also defined $J_{u,n}=J\mathcal{J}_{n}\left(A_{u}\right)$ as the renormalized hopping along the $u=x,y$ direction.
On top of this, if we assume moderate field amplitudes such that the contribution from the first Bessel function is enough to describe the dynamics, the mass term simplifies to:
 \begin{equation}
     h_{1}\left(\mathbf{k},\varphi\right) = \frac{16J_{x,1}J_{y,1}}{\omega}\sin\left(k_{x}\right)\cos\left(k_{y}\right)\sin\left(\varphi\right) . \label{eq:MassTerm}
 \end{equation}
Then, the eigenvalues of the stroboscopic Hamiltonian are:
 \begin{align}
     E_{\pm}\left(\mathbf{k}\right) =& \pm2\left[J_{x,0}^{2}\cos^{2}\left(k_{x}\right)+J_{y,0}^{2}\sin^{2}\left(k_{y}\right)\right. \nonumber\\
     &\left.+h_{1}\left(\mathbf{k},\varphi\right)^{2}\right]^{1/2} ,
     \label{eq:Energies-Square1}
 \end{align}
which show that the driving can introduce spatial anisotropy in the hopping, but more importantly, that the mass term adds an additional contribution to the eigenvalues which is maximized at the original Dirac points $\mathbf{k}= (\pm \pi/2,0)$, opening the $0$-gap.\\
Figure~\ref{fig:Square-Chern}, panel $(a)$, shows the exact quasienergies for circular polarization in the high frequency regime, demonstrating the opening of the $0$-gap.
The agreement with Eq.~\eqref{eq:Energies-Square1} is excellent, as far as the frequency is the dominant energy scale and the quasienergies in different sidebands are widely spaced.

To study the topology of the dynamically induced gap in the high frequency regime we can apply the tools of static systems to the stroboscopic Hamiltonian~\cite{Ryu_2010,chiu_classification_2016}. Then, as the system lacks any of the relevant symmetries for topology and is two-dimensional, we can conclude that its topology is characterized by a Chern number.
Its calculation can be carried out analytically~\cite{sticlet_geometrical_2012}, and we find:
 \begin{equation}
     c_{\pm} = \mp \text{sgn}\left[J_{x,0}J_{y,0}\right]\text{sgn}\left[J_{x,1}J_{y,1}\sin\left(\varphi\right)\right] .
 \end{equation}
In Fig.~\ref{fig:Square-Chern}, panel $(b)$, we show the phase diagram as a function of the field amplitude along each axis.
One can clearly see that the topological changes are produced by the zeroes of the Bessel functions $\mathcal{J}_0(A_{x,y})$ and $\mathcal{J}_1(A_{x,y})$.\\
This mechanism to create a Chern topological phase in a semi-metal using a high-frequency AC field is analogous to the one reported in previous studies~\cite{Oka-2009,gomez-leon_engineering_2014}. However, the phase diagram and its dependence on the external field parameters is particular to this lattice.

\emph{Importantly, notice that in order to find the mass term in Eq.~\eqref{eq:MassTerm}, we have not made a distinction between rotating or counter-rotating terms, because at high frequency both are expected to contribute equally.}
This will be of crucial importance in the next section.
\subsection{Derivation of the RWA Hamiltonian}
We now ask about the effect of lowering the frequency and the possibility to obtain a RWA Hamiltonian that captures the the topological changes due to resonances.

First, note that the Peierls substitution in Eq.~\eqref{eq:Undriven-Square-H}, in combination with the Jacobi-Anger expansion, already produces the required interaction picture Hamiltonian of Fig.~\ref{fig:Flow1}. This would not be the case if the coupling to the field was performed in a different gauge~\cite{Ruocco2017}, where one would need to perform the gauge transformation $R_1(t)$.\\
Next, we assume moderate field amplitudes such that the full Hamiltonian can be truncated to the contributions proportional to the zeroth and the first Bessel functions only~\footnote{We have checked numerically that this truncated Hamiltonian to the first two Bessel functions perfectly reproduces the exact quasienergies of the full model, for the range of parameters of interest.}.

In contrast with the SSH chain, now we have two choices for the static band structure of the approximate Hamiltonian: 
we can consider the static part $H^{(0)}_\mathbf{k}$, which describes a semi-metallic phase with renormalized hopping, such that case the full time-dependent Hamiltonian is:
\begin{equation}
    H_\mathbf{k}(t) \approx H_\mathbf{k}^{(0)}+H_\mathbf{k}^{(+1)} e^{i \omega t}+H_\mathbf{k}^{(-1)} e^{-i \omega t} , \label{eq:Square-TimeDep1}
\end{equation}
or we can consider $\bar{H}_\mathbf{k}$ from Eq.~\eqref{eq:Square-HF1}, which describes the insulating phase dynamically produced by the first order correction in $\omega^{-1}$.

Strictly speaking, the correct choice is the first one. This is to be expected, and we have checked that the numerical solution of Eq.~\eqref{eq:Square-TimeDep1} perfectly agrees with the numerical solution of the full time-dependent Hamiltonian, for the case of moderate field amplitudes.
However, it turns out that this choice is not compatible with the successive RWA.
This is because \emph{neglecting counter-rotating terms to capture the resonance is only correct near the resonance, but not at high frequency, where rotating and counter-rotating terms contribute similar amounts.}
In particular, one finds that applying the RWA to Eq.~\eqref{eq:Square-TimeDep1} results in an incorrect mass term at high frequency, and crucially, does not open the $0$-gap.
For this reason, the first choice gives an erroneous band structure where the $0$-gap remains closed, because it is generated from the rotating terms only.
\begin{figure}
    \centering
    \includegraphics[width=0.85\columnwidth]{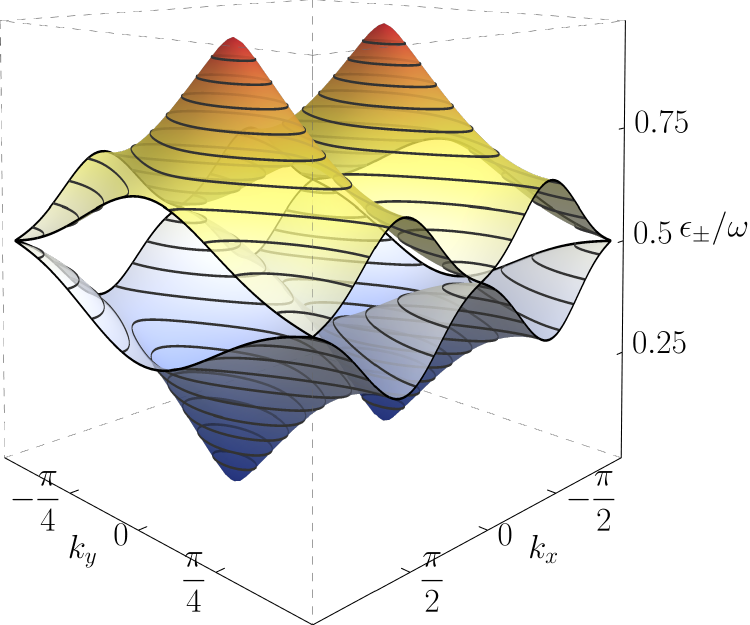}
    \caption{Quasienergy spectrum centered at the $\pi$-gap for PBC and the frequency given by the resonance condition from Eq.~\eqref{eq:Square-Resonance1}. The $0$-gap remains open due to the dynamical gap, but the $\pi$-gap exactly closes at $\mathbf{k}=( 0 , \pm \pi/2)$, producing a topological phase transition that changes the Chern number. Parameters: $A_{x}=0.5$, $A_{y}=0.5$ and $\varphi = \pi/2$.}
    \label{fig:Square-PBC-PiGap}
\end{figure}
This can be fixed using more sophisticated theoretical methods~\cite{lu_effects_2012,vogl_flow_2019,thomson_flow_2020}, but in order to avoid the use of heavier numerical calculations, let us justify the range of validity of the second option.\\

If we consider the static part to be given by $\bar{H}_{\mathbf{k}}$ instead of $H_\mathbf{k}^{(0)}$, the time-dependent Hamiltonian becomes (cf. Eq.~\eqref{eq:Square-TimeDep1}):
\begin{equation}
    H_\mathbf{k}(t) \approx \bar{H}_{\mathbf{k}} + H_{\mathbf{k}}^{(+1)} e^{i \omega t} + H_{\mathbf{k}}^{(-1)} e^{-i \omega t} . \label{eq:Square-TimeDep2}
\end{equation}
As previously mentioned, there is a redundancy in $H_\mathbf{k}(t)$, because the effect of the rotating and counter-rotating terms at high frequency is already included in $\bar{H}_{\mathbf{k}}$. 
However, we can see that the effect of the driving is quite different in resonant and off-resonant states.

In particular, the high frequency correction strongly modifies the quasienergy band structure near the Dirac points at the $0$-gap, leaving the rest of the band structure mostly unaffected (cf. Fig.~\ref{fig:Schematic-Square}, panel $(b)$, and Fig.~\ref{fig:Square-Chern}, panel $(a)$. 
Also, in the Appendix~\ref{sec:Appendix_Square}, Fig.~\ref{fig:AppendixFig3}, we plot the overlap of the two band structures for direct comparison). In contrast, the states that are resonant mainly modify the spectrum around the $\pi$-gap.
This means that if the frequency is larger than the gap opened by the mass term, but smaller than the bandwidth, we can separate the off-resonant effect for states near the Dirac points ($0$-gap opening), from the resonance effect at higher energy states, away from the Dirac points ($\pi$-gap opening).
See Fig.~\ref{fig:Schematic-Square}, panel $(c)$ for a schematic description of the energy scales involved.

Therefore, we can conclude that the approximation will be more accurate for systems with a large bandwidth, and with drive frequencies $\omega$, larger than the dynamical gap:
\begin{equation}
    4\sqrt{J_{x,0}^2 + J_{y,0}^2} \gg \omega \gg 16 J_{x,1} J_{y,1} / \omega .
\end{equation}
Under this condition, the physics of states at resonance and off-resonance is effectively decoupled and Eq.~\eqref{eq:Square-TimeDep2} provides a good approximation.
Importantly, notice that as the dynamical gap size depends on the field amplitude $A_u$, the approximation at a given frequency can usually be improved by lowering the amplitude. The reason is that a smaller amplitude reduces the size of the region of states affected by the gap opening and then, enlarges the energy separation between resonant and off-resonant states.\\

Now we perform a similar analysis as for SSH chain in the previous section.
First, we transform the time-dependent Hamiltonian from Eq.~\eqref{eq:Square-TimeDep2} to the basis of eigenstates of $\bar{H}_\mathbf{k}$:
\begin{equation}
    \hat{H}_\mathbf{k}(t) = D(\mathbf{k}) + \hat{V} \left( \mathbf{k}, t \right) ,
\end{equation}
with
\begin{equation}
    \hat{V} ( \mathbf{k}, t) = \Lambda^\dagger (\mathbf{k}) \left[ H_{\mathbf{k}}^{(+1)} e^{i \omega t} + H_{\mathbf{k}}^{(-1)} e^{-i \omega t} \right] \Lambda (\mathbf{k})
\end{equation}
and $D(\mathbf{k})$ a diagonal matrix with entries given by the energies in Eq.~\eqref{eq:Energies-Square1}.
Next, we isolate the rotating terms, which requires to keep only the term proportional to $\sigma_{-}$ from $H_{\mathbf{k}}^{(+1)}$, and the term proportional to $\sigma_{+}$ from $H_{\mathbf{k}}^{(-1)}$.
With this, we arrive to the final RWA Hamiltonian:
\begin{equation}
    H_\text{RWA}\left(\mathbf{k}, t\right) = \left(\begin{array}{cc}
E_{+}(\mathbf{k}) & \hat{V}(\mathbf{k})^{1,2}e^{-i\omega t}\\
\hat{V}(\mathbf{k})^{2,1}e^{i\omega t} & E_{-}(\mathbf{k})
\end{array}\right) . \label{eq:Square-RWA1}
\end{equation}
The solution to the time-dependent Schrödinger equation is straightforward to obtain by means of a transformation $R(t)=e^{-i\frac{\omega}{2} t \sigma_z}$, which results in the following rotating frame Hamiltonian:
\begin{equation}
    \tilde{H}_\mathbf{k} = \left(\begin{array}{cc}
E_{+}(\mathbf{k})-\frac{\omega}{2} & \Gamma (\mathbf{k})\\
\Gamma (\mathbf{k})^* & E_{-}(\mathbf{k})+\frac{\omega}{2}
\end{array}\right) , \label{eq:Square-RWA2}
\end{equation}
with eigenvalues
\begin{equation}
    \tilde{E}_{\pm}(\mathbf{k}) = \pm\sqrt{\left[ E_{+}(\mathbf{k})-\frac{\omega}{2} \right]^{2} + |\Gamma(\mathbf{k})|^2} , \label{eq:Square-Eigenvalues1}
\end{equation}
being $\Gamma (\mathbf{k}) \equiv \hat{V}(\mathbf{k})^{1,2} = [ \hat{V}(\mathbf{k})^{2,1} ]^*$.
\begin{figure*}
    \centering
    \includegraphics[width=1\textwidth]{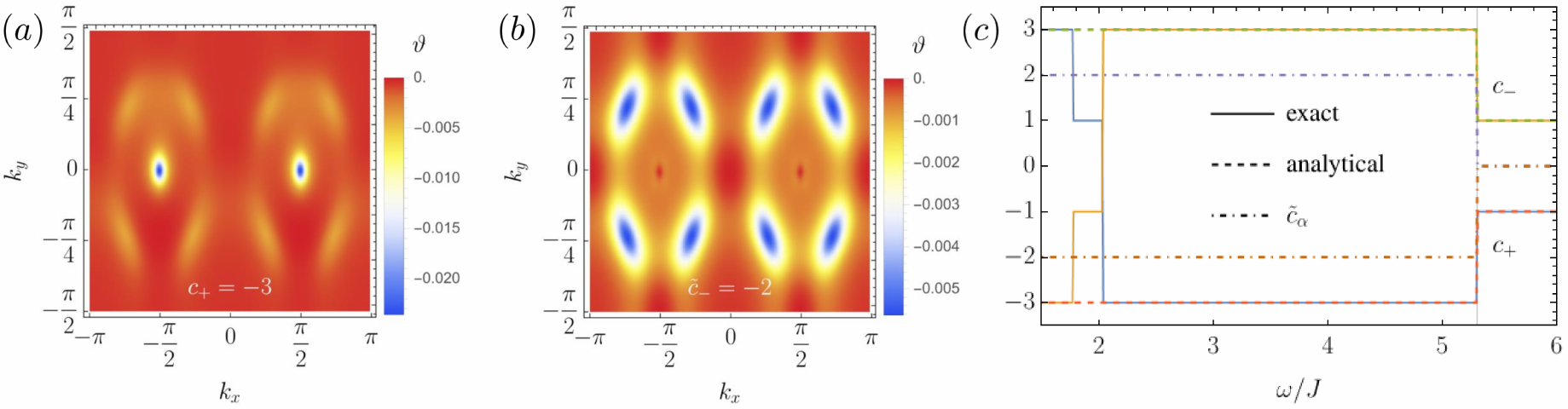}
    \caption{$(a)$ Berry flux distribution of the upper quasienergy band for a resonant frequency configuration, $\omega=3.5J$, with Chern number $c_{+}=-3$. $(b)$ Berry flux for the rotating frame Hamiltonian for identical parameters. $(c)$ Chern number as a function of frequency for identical conditions as panel $(a)$. Parameters: $\varphi=\pi/2$ and $A_{x}=A_{y}=0.5$.}
    \label{fig:Berry-Flux1}
\end{figure*}
Formally, this solution is similar to that of the SSH chain in Eq.~\eqref{eq:eigenvalues2}, and we expect a topological phase transition due to a resonance if the gap exactly closes at particular points of the FBZ.

The requirements for a gap closure are obtained from Eq.~\eqref{eq:Square-Eigenvalues1}, and correspond to the resonance condition, $\omega = 2E_{+}(\mathbf{k})$, and to the exact degeneracy condition, $\Gamma(\mathbf{k}) = 0$. 
In the Appendix~\ref{sec:Appendix_RWA} is shown a general method to solve for them and it is demonstrated that, in the present case, the $\pi$-gap closes at $\mathbf{k} = ( 0 , \pm \pi/2)$.
At these points the mass term identically vanishes and the resonance condition reduces to:
\begin{equation}
    \omega=4\sqrt{J_{x,0}^{2}+J_{0,y}^{2}} . \label{eq:Square-Resonance1}
\end{equation}
In Fig.~\ref{fig:Square-PBC-PiGap} we plot the quasienergy spectrum centered at the $\pi$-gap, for a frequency that fulfills the resonance condition in Eq.~\eqref{eq:Square-Resonance1}. It shows that the $\pi$-gap exactly closes at $\mathbf{k}=( 0 , \pm \pi/2)$, as predicted above.
Also, one can see that the $0$-gap remains open due to the off-resonant contributions, confirming the energy scale separation between resonant and off-resonant degrees of freedom.
\subsection{Topology induced by resonances}
Once we have confirmed that the quasienergy band structure is correctly captured by the RWA effective Hamiltonian, we move on to analyze the topological properties of the bands.
In particular, as we are interested in changes in the topology as the frequency is lowered, we focus on the effect of the resonance in the Chern number. 
For that, let us first study the Berry flux distribution. It is an interesting quantity because it can be measured experimentally~\cite{wintersperger_realization_2020,Sokhen2023} and because its integral over the FBZ is proportional to the Chern number.

For PBC, one can calculate the Berry flux distribution numerically, by discretizing the FBZ, finding the Berry flux along each plaquette and plotting its value over all plaquettes.
The Berry flux piercing a single plaquette is just the angle rotated by an eigenstate during parallel transport in momentum space~\cite{Fukui2005}:
\begin{align}
    \vartheta_\alpha (\mathbf{k}) =& -\text{Im} \left\{ \log \left[ \langle \Phi_\alpha(k_x,k_y) | \Phi_\alpha(k_x+\delta k_x,k_y) \rangle \right. \right. \nonumber \\
    & \langle \Phi_\alpha(k_x + \delta k_x,k_y) | \Phi_\alpha(k_x+\delta k_x, k_y+\delta k_y) \rangle \nonumber \\
    & \langle \Phi_\alpha(k_x + \delta k_x, k_y + \delta k_y) | \Phi_\alpha(k_x, k_y+\delta k_y) \rangle \nonumber \\
    & \left. \left. \langle \Phi_\alpha(k_x, k_y+\delta k_y) | \Phi_\alpha(k_x,k_y) \rangle \right] \right\} .
\end{align}
Then, the Chern number can be obtained by calculating the total flux:
\begin{equation}
    c_\alpha=\frac{1}{2\pi}\sum_{\mathbf{k}\in\text{FBZ}} \vartheta(\mathbf{k})_\alpha . \label{eq:Chern1}
\end{equation}

As a check, we first focus on the high frequency regime. The calculation of the exact Berry flux for the quasienergy bands results in the plot shown in Fig.~\ref{fig:Square-Chern}, panel $(c)$.
As expected, the Berry flux is mostly located around the Dirac points and perfectly agrees with the flux obtained from the stroboscopic effective Hamiltonian $\bar{H}_\mathbf{k}$, in Eq.~\eqref{eq:Square-HF1}.
The Chern number resulting from the sum over all plaquettes perfectly predicts the presence of edge states in the $0$-gap.
Therefore, we can conclude that the high frequency analysis for the $0$-gap topology using Eq.~\eqref{eq:Square-HF1} nicely agrees with the exact numerical results, and the Berry flux distribution is faithfully reproduced.\\

As the frequency is lowered, the picture of stroboscopic dynamics breaks down and the Berry flux changes. However, it can still be calculated numerically for the exact Floquet states.
This gives us some information about the topology, but as previously discussed, the number of edge states is not generally proportional to the value of this Chern number~\cite{gomez-leon_engineering_2014}.\\
In particular, let us consider a situation where the driving frequency is below the resonance indicated by Eq.~\eqref{eq:Square-Resonance1}. The exact Berry flux distribution for the Floquet band is shown in Fig.~\ref{fig:Berry-Flux1}, panel $(a)$.
It shows that it still is large near the original position of the Dirac points. However, there is an additional contribution distributed over a much wider area around them.
The Chern number calculated by summing the Berry flux over all plaquettes gives in this case $c_+ = -3$, which indicates that the additional contribution to the flux, due to its large extension, changes the Chern number in two units.\\
As $c_\pm \neq 0$, we can conclude that the phase including a resonance is not anomalous in this case. However, this is not important, as we are interested in the possibility to separate the topology of Floquet phases in general, not only for the anomalous case.\\

Now let us check if our analytical solution for the Floquet states can predict the number of edge states at each gap and their corresponding invariants, before we study the spectrum for OBC.
As in previous sections, the analytical expression for the Floquet states of the zero sideband can be written as
\begin{equation}
    |\Phi_{\pm}\left(\mathbf{k},t\right)\rangle = \Lambda\left(\mathbf{k}\right) e^{-i\frac{\omega}{2}t(\sigma_{z}\mp 1)}|\phi_{\pm}\left(\mathbf{k}\right)\rangle . \label{eq:Square-Analytical1}
\end{equation}
In general, the calculation of the Berry flux using Eq.~\eqref{eq:Square-Analytical1} gives an excellent agreement with the exact case. Importantly, for a correct comparison one must remember that the band ordering for the eigenvalues of the rotating frame Hamiltonian and for the quasienergies is reversed (cf. Eq.\eqref{eq:TLS-Eigen} and Eq.~\eqref{eq:quasienergies2}). Hence, numerical calculations for the quasienergy band $\alpha$ must be compared with the $-\alpha$ contributions of the analytical solution.\\

At high frequency we perfectly reproduce the distribution of Fig.~\ref{fig:Square-Chern}, panel $(c)$.
Interestingly, in Fig.~\ref{fig:Berry-Flux1}, panel $(b)$, we plot the Berry flux distribution of the eigenstates of the rotating frame Hamiltonian, $|\phi_{\pm}\left(\mathbf{k}\right)\rangle$ for a resonant configuration.
It turns out that if we remove this Berry phase distribution from the total one, which is shown in Fig.~\ref{fig:Berry-Flux1}, panel $(a)$, we reproduce the Berry flux of Fig.~\ref{fig:Square-Chern}, panel $(c)$, for the high frequency regime~\footnote{For the comparison, note the difference in scale between panels $(a)$ and $(b)$ of Fig.~\ref{fig:Berry-Flux1}}.\\
This indicates that the total Berry flux can be expressed in terms of two contributions, one coming from the renormalized bands at high frequency and one coming from the contribution of the resonance. \emph{What does this imply for the Chern number?}

Due to the general form of the Floquet states in Eq.~\eqref{eq:Square-Analytical1}, the Chern number can be separated in two contributions: one from the rotating frame Hamiltonian, $\tilde{c}_{\alpha}$, and one from the matrix $\Lambda(\mathbf{k})$ that describes the renormalized bands, $\bar{c}_{\alpha}$.
Hence, we can write for this Floquet Chern insulator in 2D:
\begin{equation}
    c_\alpha = \bar{c}_{-\alpha} + \tilde{c}_{-\alpha} , \label{eq:ChernNumbersEq}
\end{equation}
where the difference in the band index is just a consequence of the reverse ordering of the quasienergies with respect to the eigenvalues of the rotating frame Hamiltonian.

In Fig.\ref{fig:Berry-Flux1}, panel $(c)$, we plot the value of the Chern number as a function of frequency. One can see that the analytical (dashed) and the exact (solid) perfectly agree at high and intermediate frequency. Furthermore, the critical frequency $\omega\simeq 5.3J$, for the topological phase transition predicted from Eq.~\eqref{eq:Square-Resonance1} perfectly reproduces the exact critical point (vertical grey line).
At even lower frequencies, of the order of $\omega=2J$, there is a disagreement, but this is to be expected due to the resonance of states near the Dirac points, which are heavily affected by the dynamical gap.

Furthermore,  at resonance the rotating frame Chern number $\tilde{c}_\alpha$ (dot-dashed) predicts a change in two units at the $\pi$-gap, in perfect agreement with the change in the total Chern number calculated from the exact Floquet states.
This indicates that the RWA approximation for the Floquet states captures the relevant changes in the topology due to resonances, and confirms that, in analogy with the one-dimensional example, it can be separated in two contributions, one for each frame of reference, or equivalently, for each unequivalent gap.\\

\begin{figure}
    \centering
    \includegraphics[width=1\columnwidth]{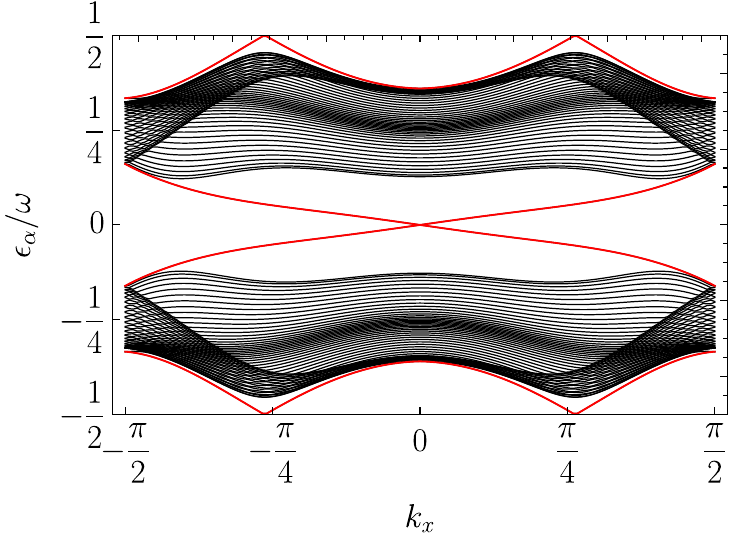}
    \caption{Quasienergy spectrum for OBC along the $y$-axis for the same parameters as those used in Fig.~\ref{fig:Berry-Flux1}. The chiral edge states are highlighted in red.}
    \label{fig:Square-OBC1}
\end{figure}

As a final check, we consider OBC and plot in Fig.~\ref{fig:Square-OBC1} the quasienergy spectrum for a ribbon with boundary along the $y$-axis. It shows that, as predicted, the $0$-gap hosts a pair of edge states (red), while simultaneously in the $\pi$-gap there are two pairs of edge states, consequence of the topological phase transition triggered by the resonance.
This shows that we have correctly predicted the number of edge states in each gap, and that they are of topological origin. Furthermore, we have checked that they also appear for OBC along the $x$-axis, indicating that the boundary-dependence of the undriven phase is irrelevant in this topological phase.\\
These results confirm that the bulk-to-edge correspondence at each gap, previously obtained for the one-dimensional case in terms of different frames of reference, is also valid in two-dimensions.
\section{Conclusions:}
In this work, we have discussed the importance of resonances in Floquet phases, and in particular, their crucial role in the existence of anomalous topology.
For that, we have provided a novel and physically intuitive approach to understand Floquet topological phases, which is valid from high to resonant frequency regime.
It shows that the topology of periodically driven systems can be systematically mapped to an effective RWA Hamiltonian, where the relation between physical processes and topology is very transparent.
On the one hand, it demonstrates that the topology of Floquet systems can be understood as the one of the $0$-gap, which is captured by the renormalized bands of the average Hamiltonian, and the one of the $\pi$-gap topology, which is related with the resonance mechanism, only activated at lower frequencies.\\
The presence of different frames of reference with topological properties confirms the importance of micromotion. However, its role is to link the different frames of reference and it does not affect the calculation of invariants.
Hence, from a topological perspective, the separation of topological invariants between different frames of reference allows us to understand the bulk-to-edge correspondence in Floquet systems from a different perspective and simplifies the calculations of invariants.

In general, our method to derive a RWA effective Hamiltonian generalizes the well-known high-frequency expansion of Floquet engineering, and by including the role of resonances, the transition between different frequency regimes can be continuously studied.
Due to this, we have obtained analytical solutions for the anomalous Floquet states of the driven SSH chain and the $\pi$-flux lattice that encode the effect of rotating, counter-rotating terms and resonances.

These results are important not only from a theoretical perspective, but also for experimental setups, as usually the description of a system with driving involves both, resonant and off-resonant states, and one cannot restrict the theory to a high-frequency expansion only.
They also allow to predict the necessary conditions for exact gap closures, based on the symmetries of the driving and lattice, and to determine the particular frequency values required for this to happen.

As one of the main findings in this work is the existence of a rotating frame Hamiltonian that captures the $\pi$-gap physics, it is interesting to understand its relation with experiments.
Its characterization could be done by spectroscopic analysis of the Floquet bands, by measuring the dependence of the $\pi$-gap with the drive frequency and the size of the avoided crossing at resonance~\cite{Gedik2013}.
Another possibility would be to measure the Berry flux distribution, as was done in~\cite{wintersperger_realization_2020,Sokhen2023}. By comparing the distributions at high frequency and at resonance, their difference should provide the Berry flux distribution that corresponds to the rotating frame Hamiltonian. Hence, this comparison could help understand the presence of anomalous topology in situations where the measured Chern number vanishes. Notice that this is different to what was proposed in~\cite{Nur2019}, where the idea is to keep track of gap closures in the spectrum, to measure the three dimensional winding number.

These results also open new horizons for future work. For example, a generalization to include higher harmonics in a unified way and to improve numerical precision can be achieved using ideas from~\cite{Benito2014}.
Also, it would be interesting to include the role of dissipation to understand the topological protection of Floquet anomalous phases to different sources of decoherence.  In addition, how these results are affected by the presence of interactions is an appealing line of research.

Finally, although this work describes two-band models, the formalism should remain valid for $N$-band systems. As the results for the $\pi$-flux lattice demonstrate, the theoretical treatment only requires an energy separation between resonant and off-resonant physics. Hence, the presence of additional bands far from the resonance can be incorporated in the standard way and would not affect the $\pi$-gap physics.
The most notable difference with the two band case would be that the rotating frame Hamiltonian becomes of dimension $N$. However, only the levels coupled by the resonance could be topologically non-trivial and its effective dimension would largely decrease.
A more complex situation would be the case of multi-chromatic drive, where several resonances can be simultaneously excited. However, this is out of the scope of this work.
\begin{acknowledgments}
AGL acknowledges P. Delplace, A. Amo, B. P\'{e}rez-Gonz\'{a}lez, G. Platero and R. Molina their useful comments and the critical reading of the manuscript.
This publication is part of the project PID2023-146531NA-I00 funded by MICIU/AEI/10.13039/501100011033 and by ERDF/EU.
AGL also acknowledges support from the European Union’s Horizon2020 research and innovation program under Grant Agreement No.899354 (SuperQuLAN) and from CSIC Interdisciplinary Thematic Platform (PTI+) on Quantum Technologies (PTI-QTEP+).
\end{acknowledgments}

\bibliographystyle{quantum}
\bibliography{AnomalousBib}
\newpage
\onecolumngrid
\appendix
\section{General analysis of RWA Hamiltonian}\label{sec:Appendix_RWA}
To find a general expression for the exact degeneracy condition of the RWA Hamiltonian, in a two-level system, we can write the eigenstates in the following generalized form:
\begin{equation}
    |v_{+}\rangle=\left(\begin{array}{c}
\cos\left(\frac{\theta}{2}\right)e^{-i\phi}\\
\sin\left(\frac{\theta}{2}\right)
\end{array}\right),\ |v_{-}\rangle=\left(\begin{array}{c}
\sin\left(\frac{\theta}{2}\right)e^{-i\phi}\\
-\cos\left(\frac{\theta}{2}\right)
\end{array}\right)
\end{equation}
Hence, the transformation matrix to the diagonal basis is:
\begin{equation}
    \Lambda=\left(\begin{array}{cc}
\cos\left(\frac{\theta}{2}\right)e^{-i\phi} & \sin\left(\frac{\theta}{2}\right)e^{-i\phi}\\
\sin\left(\frac{\theta}{2}\right) & -\cos\left(\frac{\theta}{2}\right)
\end{array}\right)
\end{equation}
Now, consider for example the $\pi$-flux lattice of this work. Taking into account that the driving term under the assumption of moderate driving is:
\begin{equation}
    H_{\mathbf{k}}^{\left(\pm1\right)}=\left(\begin{array}{cc}
0 & V_{\pm}^{1,2}\\
V_{\pm}^{2,1} & 0
\end{array}\right)
\end{equation}
with 
\begin{equation}
V_{\pm}^{1,2}=\pm2\left[iJ_{x,1}\sin\left(k_{x}\right)-J_{y,1}e^{\pm i\varphi}\cos\left(k_{y}\right)\right]
\end{equation}
and
\begin{equation}
V_{\pm}^{2,1}=\pm2\left[iJ_{x,1}\sin\left(k_{x}\right)+J_{y,1}e^{\pm i\varphi}\cos\left(k_{y}\right)\right],
\end{equation}
the transformation to the diagonal basis gives:
\begin{equation}
    \Lambda^{\dagger}H_{\mathbf{k}}^{\left(\pm1\right)}\Lambda=\frac{1}{2}\left(\begin{array}{cc}
\sin\left(\theta\right)\left(e^{-i\varphi}V_{\pm}^{2,1}+e^{i\varphi}V_{\pm}^{1,2}\right) & e^{-i\varphi}V_{\pm}^{2,1}\left[1-\cos\left(\theta\right)\right]-e^{i\varphi}V_{\pm}^{1,2}\left[1+\cos\left(\theta\right)\right]\\
e^{i\varphi}V_{\pm}^{1,2}\left[1-\cos\left(\theta\right)\right]-e^{-i\varphi}V_{\pm}^{2,1}\left[1+\cos\left(\theta\right)\right] & -\sin\left(\theta\right)\left(e^{-i\varphi}V_{\pm}^{2,1}+e^{i\varphi}V_{\pm}^{1,2}\right)
\end{array}\right)
\end{equation}
From this expression, we can extract the rotating and counter-rotating terms:
\begin{equation}
    \Gamma\left(\mathbf{k}\right)	=\frac{1}{2}e^{-i\varphi}V_{-}^{2,1}\left[1-\cos\left(\theta\right)\right]-\frac{1}{2}e^{i\varphi}V_{-}^{1,2}\left[1+\cos\left(\theta\right)\right]
	=\frac{1}{2}e^{i\varphi}V_{+}^{1,2}\left[1-\cos\left(\theta\right)\right]-\frac{1}{2}e^{-i\varphi}V_{+}^{2,1}\left[1+\cos\left(\theta\right)\right]
 \end{equation}
Hence, the condition for exact degeneracy is $\left|\Gamma\left(\mathbf{k}\right)\right|^{2}=0$.
From that, we can already see that, for example, the degeneracy will happen if $V_{\pm}^{1,2}=V_{\pm}^{2,1}=0$. Hence, as their $\mathbf{k}$-dependence goes as $\sin(k_x)$ and $\cos(k_y)$, we can conclude that the gap will exactly close at $\mathbf{k}=\pi( n , m+1/2 )$ . Notice that this is completely independent of the field values, which indicates that it is purely an effect from the symmetries of the coupling term.
\section{Additional results for the \texorpdfstring{$\pi$}{TEXT}-flux lattice}\label{sec:Appendix_Square}
\begin{figure}
    \centering
    \includegraphics[width=0.4\textwidth]{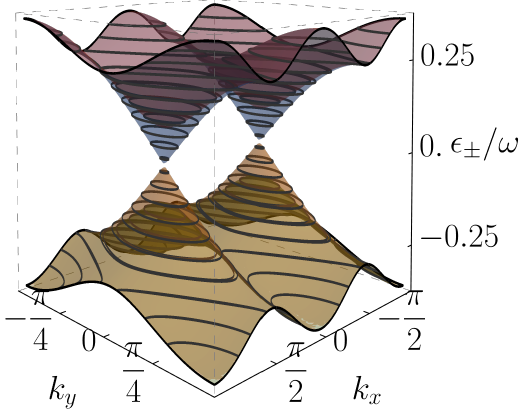}
    \caption{Comparison between the band structure before and after the drive is applied. In blue and yellow we can see the original undriven bands with their Dirac cones. Underneath are plotted in red and green the driven bands. We have considered $\omega=6J$, $A_{x,y}$=1 and $\varphi=\pi/2$.}
    \label{fig:AppendixFig3}
\end{figure}

In the derivation of the RWA Hamiltonian for the two-dimensional semi-metal we have assumed that the effect of the driving can be separated into off-resonant and resonant states. For this to be true, we need that the dynamical gap opened by the high frequency contribution is small, compared with the energy splitting of the states resonantly coupled to the field. This is possible because the renormalization of the band due to off-resonant corrections dominates close to the Dirac points. This is shown in Fig.~\ref{fig:AppendixFig3} where we compare the band before and after the high frequency field is applied. It can be seen that near the Dirac points the bands change considerably, but beyond a certain energy, the bands remain almost identical. Those are the states can be resonantly coupled by the drive in our approximation.
\end{document}